\documentclass[a4paper,USenglish,cleveref, autoref, thm-restate]{lipics-v2021}
\hideLIPIcs  %uncomment to remove references to LIPIcs series (logo, DOI, ...), e.g. when preparing a pre-final version to be uploaded to arXiv or another public repository

\EventEditors{Rotem Oshman}
\EventNoEds{1}
\EventLongTitle{37th International Symposium on Distributed Computing (DISC 2023)}
\EventShortTitle{DISC 2023}
\EventAcronym{DISC}
\EventYear{2023}
\EventDate{October 10-12, 2023}
\EventLocation{L'Aquila, Italy}
\EventLogo{}
\SeriesVolume{281}
\ArticleNo{23}

\usepackage{graphicx}
\usepackage{algorithm}
\usepackage[noend]{algpseudocode}
\usepackage{amsmath}
\usepackage{multirow}
\usepackage{graphicx}
\usepackage{xcolor}
\usepackage{amsfonts}
\usepackage{pifont}
\usepackage{xspace}
\usepackage{mathtools}
\usepackage{lineno}
\usepackage{thmtools}
\usepackage{relsize}
\usepackage{hhline}% http://ctan.org/pkg/hhline
\usepackage{etoolbox}
\usepackage{thmtools}
\usepackage{array}
\usepackage[most]{tcolorbox}
\usepackage{changepage}
\usepackage{algpseudocode}
\usepackage{hhline}
\usepackage{etoolbox}
\usepackage{relsize}
\usepackage{bm}
\usepackage{scalefnt}

\newcommand{\alglinenoNew}[1]{\newcounter{ALG@line@#1}}
\newcommand{\alglinenoPop}[1]{\setcounter{ALG@line}{\value{ALG@line@#1}}}
\newcommand{\alglinenoPush}[1]{\setcounter{ALG@line@#1}{\value{ALG@line}}}

\newcommand{\ia}{\textit{i}}
\newcommand{\ib}{\textit{ii}}
\newcommand{\ic}{\textit{iii}}

\Crefname{ALC@unique}{Line}{Lines}

\newcommand{\mypara}[1]{\smallskip\noindent\textbf{#1.}}

\newcommand{\sys}{Cordial Miners\xspace}

\newcommand{\com}[1]{}

\algdef{SE}[Receiving]{Receiving}{EndReceiving}[1]{\textbf{upon
		receiving}\ #1\ \algorithmicdo}{\algorithmicend\ \textbf{}}%
\algtext*{EndReceiving}

\algdef{SE}[Upon]{Upon}{EndUpon}[1]{\textbf{upon}\ #1\ \algorithmicdo}{\algorithmicend\ \textbf{}}%
\algtext*{EndUpon}

\newcommand\StateX{\Statex\hspace{\algorithmicindent}}
\newcommand\StateXX{\StateX\hspace{\algorithmicindent}}
\algrenewcommand\textproc{}% Used to be \textsc

\newcommand{\temph}[1]{\textbf{#1}}

\makeatletter \let\sv@thm\@thm \def\@thm{\let\indent\relax\sv@thm} \makeatother

\newcommand{\calN}{\mathbb{N}}
\newcommand{\calA}{\mathcal{A}}

\newcommand{\calE}{\mathcal{E}}
\newcommand{\calB}{\mathcal{B}}

\crefname{table}{table}{tables}
\crefname{table}{Table}{Tables}
\crefname{algocf}{alg.}{algs.}
\crefname{algocf}{Alg.}{Algs.}
\crefname{figure}{Fig.}{Figs.}
\crefname{figure}{fig.}{figs.}
\crefname{claim}{claim}{claims}
\crefname{claim}{Claim}{Claims}
\crefformat{chapter}{\S#2#1#3}
\crefmultiformat{chapter}{\S\S#2#1#3}{and~#2#1#3}{, #2#1#3}{, and~#2#1#3}

\crefformat{section}{\S#2#1#3}
\crefmultiformat{section}{\S\S#2#1#3}{and~#2#1#3}{, #2#1#3}{, and~#2#1#3}

\title{\sys: Fast and Efficient Consensus for Every Eventuality} %TODO Please add

\titlerunning{\sys} %TODO optional, please use if title is longer than one line

\author{Idit Keidar}{Technion}{}{}{}

\author{Oded Naor}{Technion and StarkWare}{}{}{}
\author{Ouri Poupko}{Ben-Gurion University}{}{}{}
\author{Ehud Shapiro}{Weizmann Institute of Science	}{}{}{}

\authorrunning{Keidar, Naor, Poupko, and Shapiro} %TODO mandatory. First: Use abbreviated first/middle names. Second (only in severe cases): Use first author plus 'et al.'

\Copyright{Idit keidar, Oded Naor, Ouri Poupko, and Ehud Shapiro } %TODO mandatory, please use full first names. LIPIcs license is "CC-BY";  http://creativecommons.org/licenses/by/3.0/


 %TODO mandatory; please add comma-separated list of keywords

\category{} %optional, e.g. invited paper

\relatedversion{Cordial Miners: Fast and Efficient Consensus for Every Eventuality} %optional, e.g. full version hosted on arXiv, HAL, or other respository/website
\relatedversiondetails
{Full Version}{https://arxiv.org/abs/2205.09174} %linktext and cite are optional

\acknowledgements{Oded Naor is grateful to the Azrieli Foundation for the award of an Azrieli Fellowship, and to the Technion Hiroshi Fujiwara Cyber-Security Research Center for providing a research grant.
Ehud Shapiro is the Incumbent of The Harry Weinrebe Professorial Chair of Computer Science and Biology at the Weizmann Institute.}%optional

\nolinenumbers %uncomment to disable line numbering

\begin{document}

\maketitle

\begin{abstract}
 Cordial Miners are a family of efficient Byzantine Atomic Broadcast protocols, with instances for asynchrony and eventual synchrony.
 They improve the latency of state-of-the-art DAG-based protocols by almost $2\times$ and achieve optimal good-case complexity of $O(n)$ by forgoing Reliable Broadcast as a building block.
 Rather, Cordial Miners use the \emph{blocklace}---a partially-ordered counterpart of  the totally-ordered blockchain data structure---to implement the three algorithmic components of consensus:  Dissemination,  equivocation-exclusion, and ordering.  
\end{abstract}

\section{Introduction}

The problem of ordering transactions in a permissioned Byzantine distributed system, also known as \emph{Byzantine Atomic Broadcast (BAB)}, has been investigated for four decades~\cite{shostak1982byzantine}, and in the last decade, has attracted renewed attention due to the emergence of cryptocurrencies.

Recently, a line of works~\cite{RN284,danezis2021narwhal,gkagol2018aleph,giridharan2022bullshark,keidar2021need,schett2021embedding} suggests ordering transactions using a distributed Directed Acyclic Graph (DAG) structure, in which each vertex contains a block of transactions as well as references to previously sent vertices.
The DAG is distributively constructed from messages of \emph{miners} running the consensus protocol.
While building the DAG structure, each miner also totally orders the vertices in its DAG locally.
That is, as the DAG is being constructed, a consensus on its ordering emerges without additional communication among the miners.

The two state-of-the-art protocols in this context are DAG-Rider~\cite{keidar2021need} and Bullshark~\cite{giridharan2022bullshark}.
DAG-Rider works in the asynchronous setting, in which the adversary controls the finite delay on message delivery between miners, and Bullshark works in the Eventual Synchrony~(ES) model, in which eventually all messages between correct miners are delivered within a known time-bound.

Both protocols use \emph{Reliable Broadcast (RB)}~\cite{bracha1987asynchronous} as a building block to disseminate vertices in the DAG.
RB ensures that Byzantine miners cannot equivocate, i.e., they cannot successfully send two conflicting vertices to the correct miners.
By using RB to exclude equivocation, the DAGs of all correct miners eventually contain the same vertices.

But using RB has costs in terms of message complexity and latency.
The well-known Bracha RB~\cite{bracha1987asynchronous} protocol entails $O(n^2)$ message complexity
for each broadcast message, where $n$ is the number of miners, and has a latency of $3$ rounds of communication.
The lower bound for RB is 2 rounds~\cite{abraham2021good}, and the message complexity lower bound is $O(n^2)$~\cite{dolev1985bounds}.
Recent RB protocols~\cite{das2021asynchronous,das2022near} improve the message complexity to $O(n)$ in some cases by using erasure codes~\cite{ben1993asynchronous}, but require between 4 to 5 rounds of communication.

DAG-Rider and Bullshark need to invoke a sequence of RB instances several times to reach a single instance of consensus.
E.g., DAG-Rider requires 6 sequential instances of RB in the expected case, making its latency between 12 to 24 rounds of communication, depending on the RB protocol it uses.
Bullshark requires between 9 to 18 rounds in the expected case in the ES model.

\begin{table*}[t] 
\resizebox{\columnwidth}{!}{%
		\begin{tabular}{|c | c |cccc|c|} 
			\hline
			\multirow{3}{*}{\textbf{Protocol}} &\multirow{3}{*}{\textbf{\begin{tabular}[c]{@{}c@{}}Reliable \\ Broadcast \\ Used\end{tabular}}} & \multicolumn{4}{c|}{\textbf{Latency}}                                                                                                                                              & \multirow{3}{*}{\textbf{\begin{tabular}[c]{@{}c@{}}Amortized \\ Message \\ Complexity\end{tabular}}} \\ 
			\cline{3-6} 
			& & \multicolumn{2}{c|}{\textbf{Eventual Synch.}} & \multicolumn{2}{c|}{\textbf{Async.}}               & \\ 
			\cline{3-6} 
			& & \multicolumn{1}{c|}{Good} & \multicolumn{1}{c|}{Expected} & \multicolumn{1}{c|}{Good} & \multicolumn{1}{c|}{Expected}& \\ 
			\hline
			\multicolumn{1}{|l|}{\multirow{2}{*}{\begin{tabular}[c]{@{}l@{}}\textbf{Cordial Miners} (this work) 	\end{tabular}}}
   & \multicolumn{1}{c|}{\multirow{2}{*}{\begin{tabular}[c]{@{}l@{}} None	\end{tabular}}} & 
   \multicolumn{1}{c|}{\multirow{2}{*}{\begin{tabular}[c]{@{}l@{}} $3$	\end{tabular}}} & \multicolumn{1}{c|}{\multirow{2}{*}{\begin{tabular}[c]{@{}l@{}} 4.5	\end{tabular}}}  &
   \multicolumn{1}{c|}{\multirow{2}{*}{\begin{tabular}[c]{@{}l@{}} $5$	\end{tabular}}} &
   \multicolumn{1}{c|}{\multirow{2}{*}{\begin{tabular}[c]{@{}l@{}} $7.5$	\end{tabular}}} &  good-case: $O(n)$ \\ 
&  & \multicolumn{1}{c|}{} & \multicolumn{1}{c|}{}  & \multicolumn{1}{c|}{} & \multicolumn{1}{c|}{} &  worst-case: $O(n^2)$ \\ 
   
			\hhline{|=|=|=|=|=|=|=|}
			
			\textbf{Bullshark} (for ES)
			& Optimal latency~\cite{abraham2021good} & \multicolumn{1}{c|}{4}  & \multicolumn{1}{c|}{\begin{tabular}[c]{@{}c@{}} $9$ \end{tabular}} & \multicolumn{1}{c|}{8} & \multicolumn{1}{c|}{12}   &  good- \& worst-case: $O(n^2)$ \\ 
			\cline{2-7} 
			\textbf{DAG-Rider} (for asynch.) & Das et al.~\cite{das2021asynchronous} & \multicolumn{1}{c|}{8}  & \multicolumn{1}{c|}{\begin{tabular}[c]{@{}c@{}}$18$ \end{tabular}} & \multicolumn{1}{c|}{16} & \multicolumn{1}{c|}{24}   &  good- \& worst-case: $O(n)$ \\ 
			\hline
			
		\end{tabular}
	}
	\caption{\textbf{Performance summary.} 
		Bullshark is for the ES model, and DAG-Rider is for the asynchronous model.
		Both protocols employ RB, which requires at least two rounds of communication of simple messages for optimal latency~\cite{abraham2021good} and $O(n^2)$ amortized message complexity, 
		or four rounds with erasure coding when using Das et al.~\cite{das2021asynchronous}
		and $O(n)$ amortized message complexity. 
	}
	\label{table:performance}
\end{table*}

It is within this context that we introduce \emph{Cordial Miners} -- a family of simple, efficient, self-contained Byzantine Atomic Broadcast~\cite{cachin2001secure} protocols that forgo RB, and present two of its instances for the models ES and asynchrony.

The ES \sys protocol reduces the expected latency from 9 rounds in today's state-of-the-art to 4.5, and the good case latency from 4 to 3.
The asynchronous version of \sys improves the expected latency from 12 rounds to 7.5, and the good case latency from 8 to 5.
This is while maintaining the same amortized quadratic message complexity in the worst case.
\sys also demonstrates better performance with $O(n)$ complexity in the good case when the actual number of Byzantine miners is $O(1)$ and the network is synchronous.
Protocols that use RB do not differ in their performance between the good and worst cases.
Tab.~\ref{table:performance} summarizes \sys' performance compared to DAG-Rider (for asynchrony) and Bullshark (for ES).

The crux of the \sys protocols is that instead of using RB to eliminate equivocation (and absorbing its rather high latency), miners cooperatively create a data structure that accommodates equivocations, termed \emph{blocklace}, which is a partially-ordered counterpart of the blockchain data structure~\cite{shapiro2023grassroots}.
When a miner wishes to disseminate a block, it simply sends it to all other miners, taking a single round of communication, instead of at least two when using reliable broadcast.

Although the blocklace may contain equivocating blocks created by Byzantine miners, they are excluded
by the ordering protocol, which is locally computed by each miner without inducing any extra communication or latency.
This is realized by the function $\tau$ that converts the partially-ordered blocklace to a totally-ordered sequence of blocks while excluding equivocations along the way.  Thus, by `complicating' the local ordering task to exclude equivocations, we forgo the extra communication rounds and latency associated with RB.

\textbf{Roadmap.}
The rest of the paper is structured as follows: \cref{section:models} describes the models and defines the problem; \Cref{section:overview} provides intuition and overview of the different components; \cref{section:blocklace} introduces the blocklace data structure; \cref{section:tau} explains the $\tau$ function that locally turns the blocklace into a totally-ordered sequence of blocks; \cref{section:protocol} describes the entire Cordial Miners protocols for the two network models; \cref{section:optimizations} presents the performance analysis; \cref{sec:related-work} is related work; and lastly, \cref{sec:futureDirections} concludes the paper.
To accommodate the space limitations some details are deferred to the appendices.
App.~\ref{appendix-section:model} describes a formal mathematical model for cordial miners, and some explanatory figures are deferred to App.~\ref{appendix-section:figures}.
Most of the proofs are deferred to App.~\ref{appendix-section:proofs}.
Potential future directions are in App.~\ref{appendix-section:futureDirection}.
Appendices \ref{appendix-section:proofs}. and \ref{appendix-section:futureDirection}. appear in the full version of this paper~\cite{keidar2022cordial}.

\section{Model and Problem Definition}\label{section:models}
We assume a set $\Pi$ of $n\ge 3$ \emph{miners} (aka agents, processes), of which at most $f < n/3$ may be \emph{faulty} (act under the control of the adversary, be `Byzantine'), and the rest are \emph{correct} (also honest or non-faulty).
Each miner is equipped with a single and unique cryptographic key-pair, with the public key known to others.  Miners can create, sign, and send messages to each other, where any message sent from one correct miner to another is eventually received. In addition, each miner can sequentially \emph{output} (aka `deliver') messages (e.g., to a local output device or storage device). 
Thus, each miner outputs a \emph{sequence} of messages.

Let $\Lambda$ denote the empty sequence; for a set $X$, $X^*$ is the set of all sequences over $X$; for sequences $x$ and $y$,  $x \preceq y$  denotes that $x$ is a prefix of $y$; $x \cdot y$ denotes the concatenation of $x$ and $y$; and $x, y$ are \emph{consistent} if  $x \preceq y$ or $y\preceq x$.

The problem we aim to solve in this paper is to devise an ordering consensus protocol that is safe and live:

\begin{definition}[Safety and Liveness of an Ordering Consensus Protocol]\label{definition:safety-liveness}
    An ordering consensus  protocol is:
    \begin{description}
	\item \textbf{Safe} if output sequences of correct miners are consistent. 
	\item \textbf{Live} if every message sent by a correct miner is eventually output by every correct miner with probability 1. 
    \end{description}
\end{definition}

Here, we aim to  devise safe and live ordering consensus protocols for models of distributed computing with two types of adaptive adversaries that can corrupt up to $f$ miners throughout the run:
First, \emph{Asynchrony}, in which the adversary controls the finite delay of every message.
Second, \emph{Eventual Synchrony (ES)}, in which there is a point in time, known as the \emph{Global Stabilization Time (GST)}. After GST, the adversary controls the delivery time of messages sent between correct miners, but they must be delivered within a known bound $\Delta$.
We further assume the adversary is computationally bounded and, therefore, cannot break cryptographic signatures.

We note that safety and liveness, combined with message uniqueness (e.g., a block in a blocklace, see next), imply the standard Byzantine Atomic Broadcast guarantees: Agreement, Integrity, Validity, and Total Order~\cite{cachin2001secure,keidar2021need}.
Hence, protocols that address the problem defined here are in fact protocols for Byzantine Atomic Broadcast.

Next, we provide an overview of the \sys protocol, including the blocklace, the dissemination of blocks, and the local ordering of the blocks to a final sequence.

\section{Cordial Miners Overview} \label{section:overview}

\begin{figure*}[t]
	\centering
	\includegraphics[width=\textwidth]{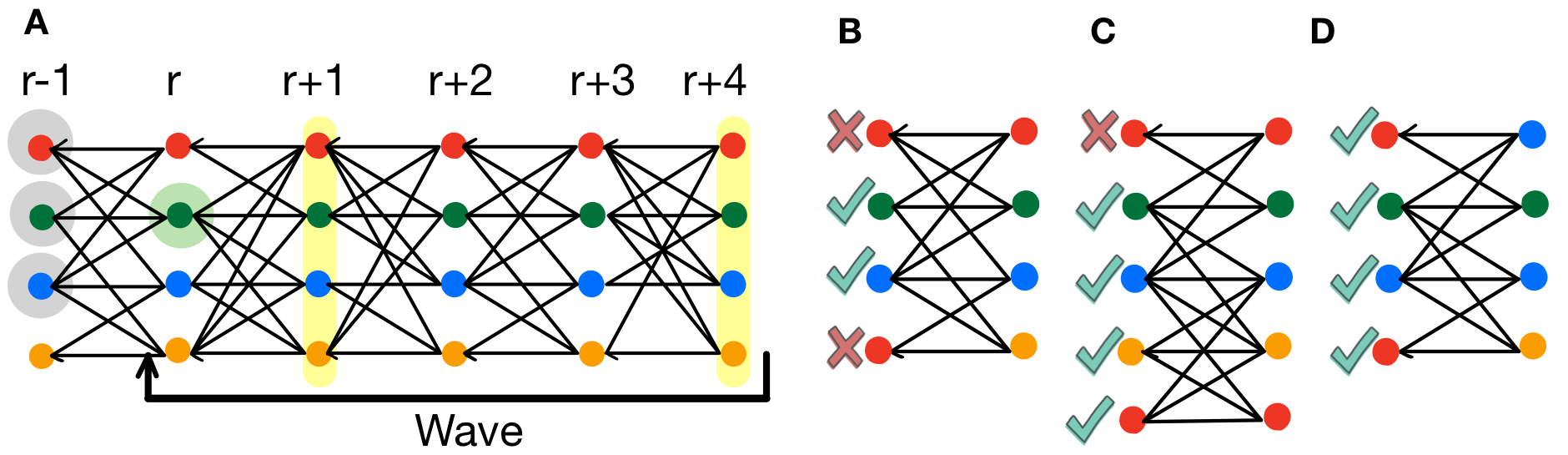}
	\caption{\smaller[0.8]\textbf{The blocklace data structure, equivocations,  approval, and ratification.} Four miners (red, green, blue, yellow). Each circle represents a block and each line a hash pointer to the left block. 
    (A) A single wave consisting of five consecutive rounds. The green block in round $r$ with the halo is the leader block. Each of the highlighted blocks in yellow in rounds $r+1$ and $r+4$ have a path to the leader block, making it a final leader block. The blocks with a gray halo are ordered by $\tau$ when the leader block becomes final.
(B) The red equivocates, with the top red block approved by the green block of the next round, the bottom red block approved by the yellow block of the next round, and the blue block of the next round, observing both equivocating red blocks, approves neither, and hence neither of the red blocks has the three approvals (including the red block itself) needed for ratification.
(C) Here the blue block of the next round observes only the bottom red block and hence approves it, which together with the yellow block and the red block itself form a supermajority, and hence the bottom red block is ratified, but not the top one. (D)  Here the blue miner equivocates in the next round, with the top blue block of the next round approving the top red block, which together with the green and red form a supermajority that ratifies it.  Similarly, the bottom blue block, the yellow block, and the red (which is not illustrated) ratify the bottom red block.  Indeed, with two equivocators (red and blue) out of four, an equivocation can be ratified.}
	\label{figure:goodcase}
\end{figure*}

In the \sys protocols, the miners jointly built  the \emph{blocklace} data structure, a partially-ordered counterpart of the totally-ordered blockchain.  A blocklace created by  four miners, each of a different color, is is illustrated in \Cref{figure:goodcase}.  The \emph{depth} of a block in a blocklace is the length of the maximal path emanating from it, and a \emph{round} of a blocklace consists of blocks of the same depth.
A set of blocks by more than $\frac{1}{2}(n+f)$ miners is termed a \emph{supermajority}; note that if $f=0$ then a supermajority is a simple majority.  Correct miners are \emph{cordial} in that they wait for round $r$ to attain a supermajority before contributing a block to round $r+1$.

\Cref{figure:goodcase}.A presents a blocklace constructed by four miners s.t. each column is a single round representing blocks from different miners, and each row in the same color consists of blocks from the same miner.
Thus, each correct miner creates a single block in each round.
Note that different miners can have different partial views of the blocklace, and the goal is to ``converge'' the order of the blocks to a consistent order for all the miners.

Each block holds a set of transactions as well as hash pointers (the edges in the DAG) to blocks of previous rounds.
When a miner observes that round $r$ has attained a supermajority (is \emph{cordial}) it creates a new block $b$ of round $r+1$ with pointers to the \emph{tips} of its blocklace up to round $r$, which are the blocks in the blocklace with no incoming edges from blocks of depth up to $r$. 
The tips must include a supermajority of blocks of round $r$, but possibly also blocks of earlier rounds not already observed by the blocks received in round $r$. (One block observes another if there is a path of pointers from one to the other.)
E.g., if the figure represents the local blocklace of the red miner, then since round $r+4$ is cordial, the red miner can create a new block $b$ in round $r+5$ with pointers to all the blocks in round $r+4$.
The miner then sends $b$ to all other miners.  The blocklace data structure is defined in \Cref{section:blocklace}.

Next, we explain how \sys use the blocklace for the three algorithmic components of consensus:  Dissemination,  equivocation-exclusion, and ordering.

\mypara{Dissemination}  In the good case, dissemination is  realized simply by each miner sending each new block to all other miners.  However, faulty miners may fail to do so, possibly intentionally, and send new blocks only to some of the miners.

The principle of \textbf{}{cordial dissemination}~\cite{shapiro2023grassroots} is: \emph{Send to others blocks you know and think they need}.    Its blocklace-based Byzantine-resilient implementation uses each block in  the blocklace as an ack/nak message:  A new block created by a correct miner $p$ points, directly or indirectly, to the blocks in $p$'s local blocklace.  It thus discloses the blocks known by $p$ at the time of its creation and, by omission, also of the blocks not yet known to $p$.
This way, a miner $q$ that receives $p$'s block can send back to $p$ any block known to $q$ and not known to $p$ according to the disclosure made by $p$'s block.
E.g., the green block in round $r+4$ serves as an ack message for all the blocks that it observes, including the red, green, and blue blocks in round $r+3$.
It also serves as a nak message for the yellow block of round $r+3$.
As an example of cordial dissemination, when the red miner sends the block it creates to the green miner in round $r+5$, it will also send to the green miner the yellow block in round $r+3$.
The dissemination protocol is formally defined in \Cref{section:protocol}.

\mypara{Equivocation exclusion} 
Two blocks $b_1, b_2$ of the same miner are \emph{equivocating} if neither observes the other, i.e., there is no path of pointers from $b_1$ to $b_2$ or from $b_2$ to $b_1$.
Since \sys do not use RB to disseminate blocks, the blocklace created by \sys may include equivocations created by Byzantine miners, which are later excluded when each miner locally orders the blocks in its blocklace to a sequence of final blocks. The \sys protocol uses supermajority approval to exclude equivocations s.t. for each set of equivocating blocks, at most, one is included in the final output.
In addition, after detecting an equivocation, correct miners ignore the Byzantine miner by not including direct pointers to their blocks.  Thus, a Byzantine miner that equivocates is eventually detected, which results in it eventually being ignored by all correct miners.  Equivocation exclusion is part of the $\tau$ ordering function which is detailed in \Cref{section:tau}.

\mypara{Ordering}
Ordering the partially-ordered blocklace can be achieved by topological sort of the DAG.
The challenge is to ensure that all correct miners exclude equivocations and order the blocks identically so that they all produce the same total order.
To this end, the blocklace is divided into \emph{waves}, each consisting of several rounds, the number of which is different for ES and asynchrony (3 and 5 rounds per wave, respectively).
E.g., \Cref{figure:goodcase}.A depicts the asynchronous version which has $5$ rounds in each wave.

For each wave, one of the miners is elected as the \emph{leader}, and if the first round of the wave has a block produced by the leader, then it is the \emph{leader block}.
The figure depicts the green block in round $r$ in the green halo as the leader block of that wave.
When a wave ends, i.e., when the last round of the wave is cordial, the leader block becomes \emph{final} if it has sufficient blocks that \emph{approve} it, namely, it is not equivocating and there is a supermajority where each block observes a supermajority that observes the leader block.
The figure shows two supermajorities, highlighted in yellow, where each block in the supermajority of round $r+4$ observes the supermajority of round $r+1$.
The supermajority in round $r+1$ observes the leader block, and the leader block is not equivocating, making it final.

A final leader block $b$ serves as the ``anchor'' of the ordering function $\tau$, which topologically sorts (while excluding equivocations) all the blocks observed by $b$ that have not been ordered yet.
Thus, each time a wave ends with a final leader block, a portion of its preceding blocklace is ordered.
In the figure, the blocks in round $r-1$ with a grey halo are ordered when the leader block in round $r$ is final since it observes them.
In case a wave ends with no final leader block,  unordered blocks will be ordered when some subsequent wave ends with a final leader block.
The full details of $\tau$ are in \Cref{section:tau}.

\section{The Blocklace}\label{section:blocklace}
A blocklace~\cite{shapiro2021multiagent} is a partially-ordered counterpart of the totally-ordered blockchain data structure: In a blocklace, each block may contain a finite set of cryptographic hash pointers to previous blocks, in contrast to one pointer (or zero for the initial/genesis block) in a blockchain. Thus, a blocklace induces a DAG in which vertices represent its blocks and edges represent the pointers among its blocks.
Next, we present the basic definitions of a blocklace, which appear as pseudocode in Alg.~\ref{alg:blocklace}.
A formal mathematical description of these definitions appears in App.~\ref{appendix-section:model}.

\subsection{Blocklace Basics}\label{subsection:blocklace-basics}

In addition to the set of miners  $\Pi$, we assume a given set of block \textbf{payloads} $\calA$, typically sets of transactions, and a cryptographic hash function \emph{hash}. A \textbf{block} consists of a payload $a \in \calA$ and a set of hash pointers to previously created blocks, signed by its creator $p$, in which case it is also
referred to as a \textbf{$p$-block} (Def. \ref{definition:block}).
A block \textbf{acknowledges} another block if it contains a hash pointer to it, and is \textbf{initial} if the set of hash pointers is empty. A \emph{blocklace} is a set of blocks (Def. \ref{definition:blocklace}). Note that $\textit{hash}$ being cryptographic implies that a blocklace that includes a cycle cannot be effectively computed, and thus a blocklace $B$ induces a DAG, with blocks as vertices in $B$ and an edge among two vertices if the first includes a hash pointer to the second.

We say that a block $b$ \textbf{observes} another block $b'$, denoted $b \succeq b'$, if there is a path from block $b$ to $b'$. If $b$ is a $p$-block in a blocklace $B$, we say that miner $p$ \textbf{observes} $b'$ \textbf{in} $B$ (Def. \ref{definition:observe}). We note that `observe' is the transitive closure of `acknowledge'. Each miner maintains a local blocklace of blocks it created and received.  With a correct miner $p$, any newly created $p$-block observes all the blocks in $p$'s local blocklace. 

The main violation a Byzantine miner $q$ can perform is an \textbf{equivocation}, by creating a pair of $q$-blocks that do not observe each other  (See Fig. \ref{figure:goodcase}.B). Such a miner $q$ is an \textbf{equivocator} (Def. \ref{definition:equivocation}). If the payloads of the two blocks are financial transactions, the equivocation may represent an attempt at double-spending.
As any $p$-block is cryptographically signed by $p$, an equivocation by $p$ is a volitional fault of $p$, to which $p$ can be held accountable.

When a block $b$ observes another block $b'$, and does not observe any equivocating block (a block $b''$ that together with $b'$ forms an equivocation), we say that $b$ \textbf{approves} $b'$ (Def. \ref{definition:approval}). Note that a block $b$ by a correct miner can observe two equivocating blocks $b', b''$, which means that $b$ approves neither $b'$ nor $b''$ (See Fig. \ref{figure:goodcase}.B). Block approval is not transitive. If $b^+$ approves $b$ and $b$ approves $b'$, yet $b^+$ also observes $b''$ (which together with $b'$ forms an equivocation), then $b^+$ does not approve $b'$. 

A miner $p$ \textbf{approves} $b'$ \textbf{in} a blocklace $B$, if $p$ has a $p$-block $b$ in $B$ that approves $b'$ (Def. \ref{definition:approval}). This holds even if $p$ has a later $p$-block $b^+$ in $B$ that observes an equivocation $b'$ and $b''$. Namely, if miner $p$ approves $b'$ in $B$ it also approves $b'$ in any $B' \supset B$.

A miner $p$ can approve both equivocating blocks $b'$ and $b''$ in a blocklace $B$, but only if $p$ is an equivocator (Obs. \ref{observation:approve-da}). An example will be if $B$ includes a $p$-block $b$ that observes $b'$ but not $b''$, and another block $b^+$ that observes $b''$ but not $b'$, which can happen only if $b$ and $b^+$ do not observe each other, namely form an equivocation (Fig. \ref{figure:goodcase}.D).

The \textbf{closure} of a block $b$, denoted $[b]$, is the set of all blocks observed by $b$. The closure of a set of blocks $B$, denoted $[B]$, is the union of the closures of the blocks in $B$. A blocklace is \textbf{closed} if it does not contain `dangling pointers' (a pointer to a block that is not in the blocklace). In other words, $B$ is closed if $B=[B]$. A block $b$ is a \textbf{tip} of a blocklace $B$ if there are no other blocks $b' \in B$ that observe $b$ (Def. \ref{definition:closure}). The \textbf{depth} (or \textbf{round}) of a block $b$ is the length of the longest path emanating from $b$.
The \textbf{depth-$d$ prefix of $B$}, denoted $B(d)$, is the set of all blocks with depth less than or equal to $d$.
The \textbf{depth-$d$ suffix of $B$}, denoted $\bar{B}(d)$, is the set of all blocks with depth greater than $d$ (Def. \ref{definition:depth}).

% \begin{algorithmic}[t]
\begin{algorithm*}[t]
    \caption{\textbf{\sys: Blocklace Utilities.} Code for miner $p $}
    \label{alg:blocklace}
   \small
    % \begin{changemargin}{0cm}{-2cm}
    \begin{algorithmic}[1] \scalefont{0.8}
        \Statex \textbf{Local variables:}
        \StateX struct $\textit{block } b$: \Comment{The structure of a block $b$ in a blocklace, Def.\ \ref{definition:block}}
        \StateXX $b.\textit{creator}$ -- the miner that created $b$ 
        \StateXX $b.\textit{payload}$ -- a set of transactions
        \StateXX $b.\textit{pointers}$ -- a possibly-empty set of hash pointers to other blocks
        \StateX \textit{blocklace} $\gets \{\}$  \Comment{The local blocklace of miner $p$}
      
		\Procedure{$\textit{create\_block}$}{$d$}: 
		\Comment Add to \emph{blocklace} a new block $b$ pointing to its tips of depth  $\le d$ 
		\State \textbf{new} $b$  \Comment{Allocate a new block structure}
		\State $b.\textit{payload} \gets \textit{payload}()$ \Comment{e.g., dequeue a payload from a queue of proposals (aka mempool)}
		\State $b.\textit{creator} \gets p$
		\State $b.\textit{pointers} \gets hash(tips)$, \text{where} \textit{tips} \text{are the tips of} \textit{blocklace\_prefix}$(d)$, \text{at most two tips per miner} \label{alg:tips} \Comment{Def. \ref{definition:closure}; two-tips limitation to prevent a Byzantine miner from flooding the blocklace before being excommunicated}

	      \State \Return $b$
		\EndProcedure
		
	\Procedure{\textit{hash}}{$b$}:
            \Return hash value of $b$ \Comment{Def.\ \ref{definition:block}}
        \EndProcedure
        
        \Procedure{$b\succeq b'$}{}: \Comment{Def.\ \ref{definition:observe}, also refereed as $b$ observes $b'$} 
        \State \Return $\exists b_1,b_2,\ldots,b_k \in \textit{blocklace}$, $k\ge 1$,  s.t.\
        $b_1 = b $, $b_k = b'$ and $\forall i \in [k-1] \colon \textit{hash}(b_{i+1}) \in b_i.\textit{pointers}$
        \EndProcedure
        
        \Procedure{\textit{closure}}{$B$}: \label{alg:SMR:closure} 
         \Return $\{b' \in \textit{blocklace} : b \in B \wedge b\succeq b'\}$ \Comment{Def.  \ref{definition:closure}. Also referred to as $[B]$. If $B=\{b\}$ is a singleton we use $[b]$ instead of $[\{b\}]$.}
        \EndProcedure
       
	\Procedure{\textit{equivocation}}{$b_1,b_2$}:  \Comment{Def.\ \ref{definition:equivocation}, Fig. \ref{figure:goodcase}.B}  \label{alg:SMR:da} 
            \State \Return 
            $b_1.\textit{creator} = b_2.\textit{creator} \wedge
            b_1 \not\succeq b_2 \wedge
            b_2 \not\succeq b_1
           $
        \EndProcedure

\Procedure{\textit{equivocator}}{$q,B$}: \label{alg:SMR:dactor}
            \Comment{Def.\ \ref{definition:equivocation}, Fig. \ref{figure:goodcase}; more faults can be added}
            \State \Return 
            $(\exists b_1,b_2 \in B :
            b_1.\textit{creator} = b_2.\textit{creator} = q 
            \wedge \textit{equivocation}(b_1,b_2))$ 
        \EndProcedure

        \Procedure{\textit{correct\_block}}{$b$}:  \Comment{See Def. \ref{definition:cordial}; other conditions can be added}
        \State \Return  
        $\{b'.\textit{creator} : \textit{hash}(b') \in b.pointers\}$ is a supermajority 
        $\wedge  \lnot\textit{equivocator}(b.\textit{creator},[b])$
        \EndProcedure

		\Procedure{\textit{approves}}{$b,b_1$}: \label{alg:SMR:approves}
            \Return $b_1 \in [b]  \wedge \forall b_2 \in [b] : \lnot$\textit{equivocation}$(b_1,b_2)$  \Comment{Def.\ \ref{definition:approval}, Fig. \ref{figure:goodcase}.C}
        \EndProcedure
          
		\Procedure{\textit{ratifies}}{$B_1,b_2$}: \Comment{Def.\ \ref{definition:ratified-block}, Fig. \ref{figure:goodcase}.C}
            \State \Return
            $
            \{b.\textit{creator} : b \in [B_1]   \wedge
              \textit{approves}(b,b_2) \}$
            is a supermajority  \label{alg:ratifies} 
          
            \EndProcedure

		\Procedure{\textit{super\_ratifies}}{$B_1,b_2$}: \Comment{Def.\ \ref{definition:ratified-block}, Fig. \ref{figure:goodcase}.A}
                \State \Return
            $
            \{b.\textit{creator} : b \in [B_1]   \wedge
              \textit{ratifies}([b],b_2) \}$ 
            is a supermajority \label{alg:super-ratifies} 
            \EndProcedure
       
        \Procedure{\textit{depth}}{$b$}:
            \State \Return max~ $\{k : \exists b' \in$ \textit{blocklace} with a path from $b$ to $b'$ of length $k$\}. \Comment{Def.~\ref{definition:depth}}
        \EndProcedure
        \Procedure{\textit{blocklace\_prefix}}{$d$}:
        \Return  $\{b \in \textit{blocklace} : \textit{depth}(b) \le d\}$  \Comment{Def.\ \ref{definition:depth}}
        \EndProcedure
        
          \Procedure{\textit{cordial\_round}}{$r$}:    \label{alg:cordial_round} 
            \State \Return 
        $\{b.\textit{creator} : b \in \textit{blocklace} \wedge \textit{depth}(b) = r\}$ is a supermajority 
        	\Comment{Def.\ \ref{definition:cordial}} \label{alg:cordial-round}
         \EndProcedure

          \Procedure{\textit{completed\_round}}{\phantom{i}}:    \label{alg:completed_round} 
            \State \Return $\max~\{ r :  \textit{cordial\_round}(r)\}$
         \EndProcedure

         %\vspace{0.5em}
        \Procedure{$\textit{last\_block}$}{$p$}: \Comment{The $p$-block with the highest round}
        \State \Return $b \in blocklace$ s.t. $b.\textit{creator}=p \wedge 
           ( \forall b'\in blocklace : b'.\textit{creator}=p \implies b'\not\succ b)$
            \EndProcedure

        \alglinenoNew{counter}
        \alglinenoPush{counter}
 
    \end{algorithmic}
      % \end{changemargin}
\end{algorithm*}
% \vspace{-2.5em}

\subsection{Blocklace Safety}
Note that as equivocation is a fault, at most $f$ miners may equivocate. Ensuring that the majority of correct miners approve a given block, requires approval from a \textbf{supermajority} of all miners, that is more than $\frac{n+f}{2}$ of the miners. A set of blocks is a \textbf{supermajority} if it includes blocks from a supermajority of miners (Def. \ref{definition:supermajority}). We show that there cannot be a supermajority approval of an equivocation (Lem. \ref{lemma:no-double-majority}).

A block $b$ \textbf{ratifies} a block $b'$ if the closure of $b$ includes a supermajority of blocks that approve $b'$. A set of blocks $B$ \textbf{super-ratifies} a block $b'$, if it includes a supermajority of blocks that ratify $b'$ (Def. \ref{definition:ratified-block} and  Fig. \ref{figure:goodcase}). 

The rounds in the blocklace are divided into \temph{waves}, such that each wave has a fixed length of $w \geq 1$, defined as the \temph{wavelength} (Def. \ref{def:leaderFunc}), and the wave consists of all the blocks in those rounds.
E.g., if the wavelength is 2, then the blocks in rounds 0 and 1 are in the first wave, and the blocks in rounds 3 and 4 are included in the second wave.
We assume the existence of a leader selection function that chooses randomly for each wave $w$ a single miner who will be the \temph{leader} of that wave.
A $p$-block $b$ is a \temph{leader block} of wave $w$ if $p$ is chosen as the leader of $w$ and the blocklace contains $b$ in the first round of $w$.
E.g., if miner $p$ is chosen as the leader of the first wave, and $p$ has a block $b$ in round $0$, then $b$ is the leader block of the first wave.
We use leader blocks as part of the ordering function $\tau$ which is detailed in \Cref{section:tau} and is used to totally order the blocklace.

Note that an equivocating leader can have several leader blocks in the same round. A leader block is \textbf{final} (Def. \ref{definition:final-leaders}) if it is super-ratified within its wave, i.e., we say that the leader block $b$ of round $r$ is final if the blocklace prefix $B(r+w-1)$ super-ratifies $b$.

The following notion of blocklace safety is the basis for the monotonicity of the blocklace ordering function $\tau$, and hence for the safety of a protocol that uses $\tau$ for blocklace ordering.
\begin{definition}[Blocklace Leader Safety]\label{definition:leader-safe}
	A blocklace $B$ is \temph{leader-safe} if every final leader block in $B$ is ratified by every subsequent leader block in $B$.
\end{definition}

A sufficient condition for blocklace leader safety is for every block in the blocklace to acknowledge blocks by at least a supermajority of miners (see Fig. \ref{figure:finality}). Such a block is a \textbf{cordial block} and a blocklace with only cordial blocks is a \textbf{cordial blocklace} (Def. \ref{definition:cordial}).
A \temph{correct block} $b$ is a $p$-block s.t. $b$ is a cordial block and $p$ does not equivocate in $[b]$ (Def~\ref{definition:correct-block}).

\begin{restatable}{proposition}{CordialBlocklaceLeaderSafety}
	\label{proposition:cordial-safety}
	A cordial blocklace is leader-safe.
\end{restatable}

\subsection{Blocklace Liveness}

Next, we discuss conditions that ensure blocklace leader liveness.

\begin{definition}[Blocklace Leader Liveness]\label{definition:leader-live}
    A blocklace $B$ is \temph{leader-live} if for every block $b \in B$ by a miner not equivocating in $B$ there is a final leader block in $B$ that observes $b$. 
\end{definition}

Given a blocklace, a set of miners $P$ is (mutually) \textbf{disseminating} if every block by a miner in $P$ is eventually observed by every miner in $P$ (Def. \ref{definition:disseminating}). We show that dissemination is unbounded, meaning that if a set of miners $P$ is disseminating in $B$  then $B$ is infinite, and in particular any suffix of $B$ has blocks from any member of $P$ (Obs. \ref{observation:infinite-dissemination}). It follows that a cordial blocklace with a non-equivocating and disseminating supermajority of miners is leader-live (Fig. \ref{figure:liveness-condition}). 

\begin{restatable}[Blocklace Leader Liveness Condition]{proposition}{BlocklaceLeaderLivenessCondition}
	\label{proposition:liveness-condition}
	If $B\subset \calB$ is a cordial blocklace with a non-equivocating and disseminating supermajority of miners, such that for every $r>0$ there is a final leader block of round $r'>r$, then $B$ is leader-live.
\end{restatable}

\section{Blocklace Ordering with \texorpdfstring{$\tau$}{tau}}\label{section:tau}

% \begin{figure}[t]
\begin{algorithm*}[t] 
    \caption{\textbf{\sys: Ordering of a Blocklace with $\boldsymbol{\tau}$} \\ pseudocode for miner $p \in \Pi$, including Algorithms \ref{alg:blocklace} \& \ref{alg:protocols}} \label{alg:tau}
%    \small
    % \begin{changemargin}{0cm}{-2cm}
    \begin{algorithmic}[1] \scalefont{0.8}
        \alglinenoPop{counter} 
	\Statex \textbf{Local Variable:}
            \StateX $\textit{outputBlocks} \gets \{\}$
	\Procedure{$\tau()$}{}: \label{alg:tau-def} \Comment{Called from \cref{alg:call-tau}} \label{alg:CMO:waveCompletion}
            \State $\tau'(\textit{last\_final\_leader}())$
	\EndProcedure \label{alg:CMO:commitrule}

        \Procedure{$\tau'$}{$b_1$}: \label{alg:CMO:da}
            \If{$b_1 \in \textit{outputBlocks} \vee b_1 = \emptyset$}
                \Return
            \EndIf
		\State  $b_2 \gets \textit{previous\_ratified\_leader}(b_1)$
            \State  $\tau'(b_2)$ \Comment{Recursive call to $\tau'$}
            \State \textbf{output} $\textit{xsort}(b_1,[b_1] \setminus [b_2])$ \Comment{Output a new equivocation-free suffix} \label{alg:CMO:Okazaki}
            \State $\textit{outputBlocks} \gets \textit{outputBlocks} \cup \textit{xsort}(b_1,[b_1] \setminus [b_2])$
        \EndProcedure
        
        \Procedure{\textit{xsort}}{$b,B$}:  \Comment{Exclude equivocations and sort} 
            \State \Return topological sort wrt $\succ$ of the set $\{b'\in B : \textit{approves}(b,b')\}$ 
        \EndProcedure
	    
        \Procedure{\textit{previous\_ratified\_leader}}{$b_1$}: \label{alg:CMO:dr1} 
            \State \Return $\textit{arg}_{b \in R} \max~  \textit{depth}(b)$
            \State where $R = \{ b \in [b_1]\setminus \{b_1\}  : b.\textit{creator} = \textit{leader}(\textit{depth}(b)) \wedge \textit{ratifies}([b_1],b)\}$ 
        \EndProcedure
        
        \Procedure{$\textit{last\_final\_leader}()$}{}: \label{alg:CMO:dr2}        \Comment{Fig. \ref{figure:finality}}
            \State \Return $\textit{arg}_{u \in U} \max~ \textit{depth}(u)$  where  
            \State $U = \{ b \in \textit{blocklace} : b.\textit{creator} = \textit{leader}(\textit{depth}(b)) \wedge \textit{final\_leader}(b)   \}$ 
        \EndProcedure

        \Procedure{\textit{final\_leader}}{$b$}: \Comment{Def. \ref{definition:final-leaders}} \label{alg:finalLeader}
        \State \Return $\textit{super-ratifies}((\textit{blocklace\_prefix}(\textit{depth}(b)+w-1), b)$
        % \oded{changed the blue line over the next line. The super-ratifies function is defined in algorithm 1.}

        \Statex \textbf{procedure} \textit{leader}$()$ (Def. \ref{def:leaderFunc}) and wavelength $w$ are defined in Alg. \ref{alg:protocols}.
         \EndProcedure
         
        \alglinenoPush{counter}	
    \end{algorithmic}
    % \end{changemargin}
\end{algorithm*}
% \vspace{-2em}

% \end{figure}

Here we present a deterministic function $\tau$ that, given a blocklace $B$, employs final leaders to topologically sort $B$ into a sequence of its blocks, respecting $\succ$.  The intention is that in a blocklace-based  ordering consensus protocol, each miner would use $\tau$ to locally convert their partially-ordered blocklace into the totally-ordered output sequence of blocks.

The section concludes with Theorem \ref{theorem:tau-conditions-safety-liveness}, which provides sufficient conditions for the safety and liveness of any blocklace-based ordering consensus protocol that employs $\tau$. The proof method is novel, in that it does not argue operationally, about events and their order in time, but rather about the properties of an infinite data structure -- the blocklace.
In the following section, we prove that the Cordial Miners protocols, which employ  Alg. \ref{alg:tau} that realizes $\tau$, satisfy these conditions, and thus establish their safety and liveness.
The operation of $\tau$ is depicted in Fig.~\ref{figure:tau}.

We show that $\tau$ is monotonic, in that if it is repeatedly called with an ever-increasing blocklace then its output is an ever-increasing sequence of blocks. This monotonicity ensures finality, as it implies that any output will not be undone by a subsequent output.  With $\tau$, final leaders are the anchors of finality in the growing chain,  each `writes history' backward till the preceding final leader.  

The following recursive ordering function $\tau$ maps a blocklace into a sequence of blocks, excluding equivocations along the way. Formally, the entire sequence is computed backward from the last super-ratified leader, afresh by each application of $\tau$.  Practically, a sequence up to a super-ratified leader is final (Prop. \ref{proposition:tau-finality}) and hence can be cached, allowing the next call to $\tau$ with a new super-ratified leader to be computed backward only till the previously-cached super-ratified leader, while
producing as output all the blocks approved by the new super-ratified leader (the approval ensures that the new fragment does not introduce equivocations) that are not observed by the previously-cached final leader.

\begin{definition}[$\tau$]\label{definition:tau}
	We assume a fixed topological sort function $\textit{xsort}(b,B)$ (exclude and sort) that 
	takes a block $b$ and a blocklace $B$, and returns a sequence consistent with $\succ$ of all the blocks in $B$ that are approved by $b$. The function $\tau: 2^{\calB} \xrightarrow{} \calB^*$ 
	is defined for a blocklace $B\subset \calB$ backward, from the last output element to the first, as follows:
	If $B$ has no final leaders then $\tau(B) :=\Lambda$ (empty sequence).
	Else, let $b$ be the last final leader in $B$. Then $\tau(B) := \tau'(b)$, where $\tau'$ is defined recursively:
	$$
	\tau'(b) := 
	\begin{cases}
		\textit{xsort}(b,[b]) \text{\ \ \ \ \   if $[b]$ has no leader ratified by $b$, else }\\
		\tau'(b') \cdot \textit{xsort}(b,[b]\setminus [b']) \text{\ \   if $b'$ is the last leader } \\
		\text{\phantom{zzzzzzzzzzzzzzzzzzzzzzzzzz}ratified by $b$ in $[b]$}
	\end{cases}
	$$
\end{definition}
Note that when $\tau'$ is called with a leader $b$, it makes a recursive call with a leader ratified by $b$, which is not necessarily super-ratified.

A pseudo-code implementation of $\tau$ is presented as Alg. \ref{alg:tau}.
The algorithm is a literal implementation of the mathematics described above:
It maintains \emph{outputBlocks} that includes the prefix of the output  $\tau$ that has already been computed.
Upon adding a new block to its blocklace (Line \ref{alg:CMO:waveCompletion}), it computes the most recent final leader $b_1$ according to Definition \ref{definition:final-leaders}, and applies $\tau$ to it, realizing the mathematical definition of $\tau$ (Def. \ref{definition:tau}), with the optimization, discussed above, that a recursive call with a block that was already output is returned.  Hence the following proposition:

\begin{proposition}[Correct implementation  of $\tau$]\label{proposition:tau-tau}
	The procedure $\tau$ in Alg. \ref{alg:tau} correctly implements the function  $\tau$ in Definition \ref{definition:tau}.
\end{proposition}

The following theorem provides a sufficient condition for the safety and liveness (Def. \ref{definition:safety-liveness}) of any blocklace ordering consensus protocol that employs $\tau$, and thus offers conditions for solving the problem defined in \Cref{section:models}:
\begin{restatable}[Sufficient Condition for the Safety and Liveness of a Blocklace-Based Ordering Consensus Protocol]{theorem}{SufficientCondition}\label{theorem:tau-conditions-safety-liveness}
Assume a given blocklace-based  consensus protocol that employs $\tau$ for ordering.  If in every run of the protocol all correct miners have in the limit the same blocklace $B$ that is leader-safe and leader-live, then the protocol is safe and live.
\end{restatable}

Next, we provide a proof outline of \Cref{theorem:tau-conditions-safety-liveness}.

\mypara{$\tau$ Safety}
A safe blocklace ensures a final leader is ratified by any subsequent leader, final or not.  Hence the following:

\begin{restatable}[Monotonicity of $\tau$]{proposition}{MonotonicityofTau}\label{proposition:tau-finality}
	Let $B$ be a cordial blocklace with a supermajority of correct miners.
	Then $\tau$ is monotonic wrt the superset relation among closed subsets of $B$, namely
	for any two closed blocklaces $B_2 \subseteq B_1 \subseteq B$,  $\tau(B_2) \preceq \tau(B_1)$.
\end{restatable}

The following proposition ensures that if there is a supermajority of correct miners, which jointly create a cordial blocklace, then the output sequences computed by any two miners based on their local blocklaces would be consistent.  This establishes the safety of $\tau$ under these conditions.

\begin{restatable}[$\tau$ Safety]{proposition}{TauSafety}\label{proposition:tau-safety}
	Let $B$ be a blocklace with a supermajority of correct miners.
	Then for every $B_1, B_2 \subseteq B$, $\tau(B_1)$ and $\tau(B_2)$ are consistent.
\end{restatable}

\mypara{$\tau$ Liveness}
While $\tau$ does not output all the blocks in its input, as blocks not observed by the
last final leader in its input are not in its output, the following observation and proposition set the conditions for $\tau$ liveness:

\begin{restatable}[$\tau$ output]{observation}{TauOutput}
	\label{observation:tau-output}
	If a $p$-block $b \in B$ by a miner $p$ not equivocating in $B$ is observed by a 
	final leader in $B$, then $b \in \tau(B)$.
\end{restatable}

\begin{restatable}[$\tau$ Liveness]{proposition}{TauLiveness}\label{proposition:tau-liveness}
	Let  $B_1 \subset B_2 \subset \ldots$ be a sequence of finite blocklaces  for which $B= \bigcup_{i\ge 1}B_i$ is a cordial leader-live blocklace.
	Then for every block $b\in B$ by a correct miner in $B$ there is an $i\ge 1$ such that $b \in \tau(B_i)$.
\end{restatable}

Thus, we conclude that the safety and liveness properties of $\tau$ carry over to Alg. \ref{alg:tau}.

Next, we prove that the two Cordial Miners consensus protocols---for eventual synchrony and asynchrony---satisfy the conditions of Theorem \ref{theorem:tau-conditions-safety-liveness}, and hence are safe and live.

\section{The Cordial Miners Protocols}\label{section:protocol}
% \begin{figure}[t]
\begin{algorithm*}[t]
	\caption{\textbf{\sys: Blocklace-Based Dissemination}\\ Code for miner $p$, including Algorithms \ref{alg:blocklace},  \ref{alg:tau} \&  \ref{alg:protocols}}	\label{alg:dissemination}
%	\small
	% \begin{changemargin}{0cm}{-1.5cm}
	\begin{algorithmic}[1]  \scalefont{0.8}
	\alglinenoPop{counter}

	\Statex \textbf{Local variables:}
	\StateX $r \gets 0$ \Comment{The current round of $p$,  see Def. \ref{definition:depth}} 
  
        \Upon{\textbf{receipt} of $b : \textit{b.pointers} \subseteq \textit{hash}(\textit{blocklace}) \wedge \textit{correct\_block}(b)$}  \Comment{Received `out of order' blocks are buffered; incorrect blocks are ignored}
	\label{alg:receive}
    	\State $\textit{blocklace} \gets \textit{blocklace}~ \cup \{ b\}$ 	
    	\State $\tau()$   \label{alg:call-tau}  \Comment{Defined in \cref{alg:tau-def}}
	    \If{ $\textit{completed\_round}() \ge r$}   \label{alg:completed-round} \Comment{Defined in \cref{alg:completed_round}, line \ref{alg:completed_round}}
        \State $\textit{es\_advance\_round}()$ \Comment{Advance round conditions for ES, no-op for asynchrony. Defined in \cref{alg:protocols}}
	    \State $b \gets \textit{create\_block}(\textit{completed\_round}())$  \label{alg:createBlock} 
	    \State $r \gets \textit{depth}(b)$ \label{alg:newDepth}  \Comment{Advance round} 
			\For{$q \in \Pi$} \Comment{Cordial Dissemination} 
		        \State  \textbf{send}  $\{b\} \cup \textit{blocklace\_prefix}(r-2) \setminus [\textit{last\_block}(q)]$ to $q$  \label{alg:send-block}

		   \EndFor  
	    \EndIf
	\EndUpon
		\alglinenoPush{counter}
	\end{algorithmic}
	
	% \end{changemargin}
\end{algorithm*}
% \vspace{-2em}
% \end{figure}

So far, we presented the blocklace and how a blocklace can be totally ordered using $\tau$.
Next, we show how miners disseminate their blocks to form a blocklace.

The shared components of the Cordial Miners protocols are specified via pseudocode in Algs. \ref{alg:blocklace} (blocklace utilities), \ref{alg:tau} (the ordering function $\tau$), and \ref{alg:dissemination} (dissemination).
Alg. \ref{alg:protocols} details the differences between the Cordial Miners protocols for ES and asynchrony.
We begin by explaining the dissemination protocol.

% \begin{figure}[t]
\begin{algorithm*}[t]
	\caption{\textbf{\sys: Specific Utilities.} Code for miner $p$. 
 %Changes between asynchrony and ES are highlighted in \textcolor{blue}{blue}.
 }
	\label{alg:protocols}
%	\small
%        \vspace{-1em}
	\begin{flushleft}
        \noindent\textbf{\ref{alg:protocols}.1 Procedures for Asynchrony}
        \end{flushleft}
%        \vspace{-1em}
      \begin{algorithmic}[1]  \scalefont{0.8}
      \alglinenoPop{counter} 
      
        \State \textcolor{black}{$w \gets 5$ \label{alg:w-5}}        
 
       \Procedure{$\textit{es\_advance\_round}()$}{}: \label{alg:timeoutA} \Comment{No-op}
        \State \Return
          % if $d \text{ mod } w = 0$ else $\bot$. 
        \EndProcedure
 
       \Procedure{\textit{leader}}{$d$}: \label{alg:leadAync} 
        \If {$d \text{ mod } w = 0$} 
        \State \textcolor{black}{\Return $q \in \Pi$ via a shared coin tossed at round $d+w-1$}
        \Else 
        \State \Return $\bot$
        \EndIf 
          % if $d \text{ mod } w = 0$ else $\bot$. 
        \EndProcedure
 
        % \State \Return $\{b' \in \textit{blocklace\_prefix}(\textit{depth}(b)+w-1) : \textit{ratifies}([b'],b)\}$ is a supermajority

		\alglinenoPush{counter}
	\end{algorithmic}
    % \end{changemargin}
%    \vspace{-1.5em}
    \begin{flushleft}
    \noindent \textbf{\ref{alg:protocols}.2 Procedures for Eventual Synchrony}
    % \oded{Udi, change the algorithm for ES to include a timeout after a round is cordial to allow for the good case performace}
    \end{flushleft}
%    \vspace{-1em}
    % \begin{changemargin}{0cm}{-2cm}
      \begin{algorithmic}[1]  \scalefont{0.8}
	\alglinenoPop{counter} 
 
		\State \textcolor{black}{$w \gets 3$} \label{alg:w-2}

       \Procedure{$\textit{es\_advance\_round}$()}{}: \label{alg:timeoutES} 
         \State \Return $\max~ r : \textit{cordial\_round}(r) ~\wedge$ \Comment{Last cordial round, \cref{alg:cordial-round}}

        \State (($r \text{ mod } w = 0 \implies$ \Comment{First round of the wave, leader is included in the round.} \label{alg:advanceCordial1}
         \State $\exists b \in \textit{blocklace}: (\textit{leader}(r) = b.\textit{creator})~ \wedge$ 
    
         \State $( (r \text{ mod } w = 1 \implies$ \Comment{Second round of the wave, round $r-1$ leader is ratified by round $r$ blocks} \label{alg:odd-round} \label{alg:advanceCordial2}
         \State $\exists b \in \textit{blocklace}: (\textit{leader}(r-1) = b.\textit{creator}~ \wedge$ 
          $\textit{ratifies}(\textit{blocklace\_prefix}(r),b)))~ \wedge$ 
           \label{alg:ratified-es}   
        
       \State $( (r \text{ mod } w = 2 \implies$ \Comment{Third round, round $r-2$ leader is super-ratified by $r$ blocks} \label{alg:second-round} \label{alg:advanceCordial3}
         \State $\exists b \in \textit{blocklace}: (\textit{leader}(r-2) = b.\textit{creator}~ \wedge$ 
          $\textit{super-ratifies}(\textit{blocklace\_prefix}(r),b)))$ 
           \label{alg:ratified2-es} 
          
        \State  $\vee~  \textit{timeout})$  \label{alg:timeout-es}    \Comment{Or timeout occurred. \textit{timeout} is measured from when round $r$ is cordial. This is $p$'s estimation of $\Delta$.}
        
        % \State \textcolor{blue}{\textbf{wait} $\textit{timeout}$ \label{alg:timeout-2delta}} \Comment{Introduced for simplicity. Can be replaced with Alg.~\ref{alg:optimization}.}
        % \State \Return
          % if $d \text{ mod } w = 0$ else $\bot$. 
        \EndProcedure

        \Procedure{\textit{leader}}{$d$}:  \label{alg:leader_es} 
        \If {$d \text{ mod } w = 0$} 
        \State \textcolor{black}{\Return $q \in \Pi$ selected deterministically}
        \Else 
        \State \Return $\bot$
        \EndIf 
        \EndProcedure
        % \label{alg:completed-es} 
        % \State \Return $\max~ r : \textit{cordial\_round}(r) ~\wedge$ \Comment{Last cordial round, \cref{alg:cordial-round}}
       
        %  \State $( (r \text{ mod } w = 1 \implies$ \Comment{Odd round, round $r-1$ leader is ratified by round $r$ blocks} \label{alg:odd-round}
        %  \State $\exists b \in \textit{blocklace}: (\textit{leader}(r-1) = b.\textit{creator}~ \wedge$ 
        %   $\textit{ratifies}(\textit{blocklace\_prefix}(r),b)))~ \wedge$ 
        %    \label{alg:ratified-es}   
        
        %   \State ($r \text{ mod } w = 0 \implies$ \Comment{Even round, round $r-2$ leader is super-ratified by round $r$ blocks}
        %      \State $\exists b \in \textit{blocklace}: (\textit{leader}(r-2) = b.\textit{creator}~ \wedge$ 
        %   $\textit{super\_ratifies}(\textit{blocklace\_prefix}(r),b))~ \wedge$
        %   \label{alg:super-ratified-es}   
        %   \State $\exists b' \in \textit{blocklace}: (\textit{leader}(r) = b'.\textit{creator}~ \wedge$ 
        %   $\textit{ratifies}([b'],b))$ %
        %    \Comment{and round $r-2$ leader is ratified by the leader of round $r$} 
        %    \label{alg:leader-ratified-es}   
        %     \label{alg:included-es}
        % \State  $\vee~  \textit{timeout})$  \label{alg:timeout-es}    \Comment{Or timeout occurred. \textit{timeout} is measured from when round $r$ is cordial. This is $p$'s estimation of the network delay $\Delta$.} 

		\alglinenoPush{counter}
  \end{algorithmic}	
	% \end{changemargin}
\end{algorithm*}
% \vspace{-2em}
% \end{figure}

\begin{table*}[t]
	
	\centering
        \resizebox{\columnwidth}{!}{%
	\begin{tabular}{ | m{9.5em} | m{12em}|  m{14.5em} |} 
		
		\hline
		\textbf{Property}  &  \textbf{Asynchrony} & \textbf{Eventual Synchrony} \\  
		\hline\hline
		
		\textbf{Wavelength $\boldsymbol{w}$:}  &  $5$ (Line \ref{alg:w-5}) &   $3$ (Line \ref{alg:w-2}) \\ 
		\hline
		\textbf{Leader Selection:} &   Retrospective via coin toss (Line \ref{alg:leadAync})  & Prospective by a known order \newline (Line \ref{alg:leader_es})  \\ 
		\hline

		\textbf{Condition for advancing round:} & None (Line \ref{alg:timeoutA}) & $\textit{timeout}$ or finality conditions \newline (Line \ref{alg:timeoutES}) \\
		\hline

		\hline
		
	\end{tabular}
        }
	\caption{\sys' differences between Eventual Synchrony and Asynchrony}
	\label{table:protocols}
\end{table*}

\subsection{Dissemination (Alg. \ref{alg:dissemination})}  
A correct block is buffered until it has no dangling pointers, and then it is received  (Line \ref{alg:receive}).
We prove that an equivocating miner eventually can only produce incorrect blocks (Def.~\ref{definition:correct-block}) and therefore is eventually excommunicated by all correct miners.
After including a received block in its local blocklace, a miner calls $\tau$ (Line \ref{alg:call-tau}), which outputs new blocks if the received block results in the blocklace having a new final leader block.

If there is a new completed round in the blocklace (Line \ref{alg:completed-round}), the miner creates a new block $b$ (Line \ref{alg:createBlock}), computes the new round (Line \ref{alg:newDepth}), and sends $b$ to its fellow miners.
The package sent to miner $q$ contains any blocks up to the previous round that $p$ knows that $q$ might not know, based on the last block received from $q$ (Line \ref{alg:send-block}).
Note that as the network is reliable, \textbf{send} is defined to be idempotent, namely to send each block to each miner at most once. 

We note that there is a tradeoff between latency and message complexity, and there is a range of possible optimizations and heuristics.
These are discussed in \Cref{appendix-section:futureDirection}. Here, we present a version of Cordial Miners protocols in which every block is communicated among every pair of correct miners in the worst case. 

\subsection{Specific utilities (Alg. \ref{alg:protocols})}
\mypara{Overview}
There are several differences between the Cordial Miners protocols for ES and asynchrony, which are specified in Alg. \ref{alg:protocols} and summarized in Tab. \ref{table:protocols}. 

First, in asynchrony, each wave consists of~$5$ rounds, and the leader block in the first round is chosen randomly using a shared coin tossed in the last round of the wave, i.e., the leader election is retrospective.
We expand on the coin below. 
In ES, each wave has~$3$ rounds, and the leader block is elected in advance using any deterministic prospective method, e.g., round robin.

The reason a wave in asynchrony is longer is to counter the adversary: If the adversary knows in advance the leader block in the first round of the wave, it can manipulate block arrival times s.t. a wave with a final leader block will never happen.
We prove that by using such coin at the last round of the wave, the adversary cannot affect the probability the the leader block is final.
In ES, a wave consists of three rounds.
We prove that this is sufficient to allow super-ratification of the leader block, making it final in case the leader is an honest miner.

Another difference is if an honest miner waits before proceeding to the next round when the current round becomes cordial.
In asynchrony, the miner proceeds immediately to the next round when it is cordial (Line \ref{alg:timeoutA}).
In ES, a miner advances to the next round after a round either if \textit{timeout} passes, or conditions for leader block finality occur (Line \ref{alg:timeoutES}).
The conditions are: if this is the first round of a wave, then the round contains the leader block (Line~\ref{alg:advanceCordial1}).
If this is the second round, then the miner advances immediately if the round has a supermajority of blocks that ratifies the leader block (Line~\ref{alg:advanceCordial2}), and lastly, if the third round of a wave has a supermajority of blocks that super-ratifies the leader block (Line~\ref{alg:advanceCordial3}).
These conditions are to prevent the adversary from ordering the messages after GST, in particular, the leader block and the blocks that super-ratify it, as the leader is known in advance.

\mypara{Algorithm walkthrough}
The \textit{leader} (Lines \ref{alg:leadAync}, \ref{alg:leader_es}) procedure, which is called as part of~$\tau$, is an implementation of Def.~\ref{def:leaderFunc}.

The Cordial Miners asynchrony protocol, for which $w=5$ (Line \ref{alg:w-5}), elects leaders retrospectively using a shared random coin. To elect the leader of round $r$, when $r \text{ mod } 5 = 0$,  all correct miners toss the coin in round $r+3$ and know in round $r+4$ the elected leader of round $r$, as follows.
We assume two \temph{shared random coin} functions: \textit{toss\_coin} and \textit{combine\_tosses}.
The function $\textit{toss\_coin}(p_s,d)$  takes the secret key $p_s$ of miner $p \in \Pi$ and a round number $d \geq 0$ as input, and produces $p$'s share of the coin of round $d$, $s_{p,d}$, as output.
If the protocol needs to compute the shared random coin for round $d$, then $s_{p,d}$ is incorporated in the payload of the $d$-depth $p$-block of every correct miner $p$.  The function $\textit{combine\_tosses}(S,d)$ takes a set  $S$ of shares  $s_{p,d}$, $d \geq 0$, for which  $|\{p : s_{p,d} \in S\}| > f+1$, and returns a miner $q \in \Pi$.
The properties of a similar function were presented in~\cite{keidar2021need}, which details how to implement such a coin as part of a distributed blocklace-like structure.

We formally define the shared coin in definition \ref{definition:shared-random-coin}. Examples of such a coin implementation using threshold signatures~\cite{boneh2001short,libert2016born,shoup2000practical} are in~\cite{cachin2005random,keidar2021need}.
The ES protocol elects leaders in a prospective manner via a fixed deterministic function, e.g., round-robin between the miners.

\subsection{Correctness Proof Outline}
The main theorem we prove is the following:
\begin{restatable}
	[Cordial Miners Protocols Safety and Liveness]{theorem}{safetyAndLiveness}\label{theorem:CM-safety-liveness}
	The protocols for  eventual synchrony and asynchrony specified in Algs. \ref{alg:blocklace}, \ref{alg:tau}, \ref{alg:dissemination}, \& \ref{alg:protocols} are safe and live (Def. \ref{definition:safety-liveness}).
\end{restatable}

We argue that in the limit the blocklaces of correct miners that participate in a run of a Cordial Miners protocol are identical, are leader-safe, and leader-live.  

A formal description of blocklace-based protocols in terms of asynchronous multiagent transition systems with faults has been carried out in reference~\cite{shapiro2021multiagent}.  Here, we employ pseudocode, presented in Algorithms \ref{alg:blocklace}, \ref{alg:tau}, \ref{alg:dissemination} \& \ref{alg:protocols}  to describe the correct behaviors of a miner in a protocol, and discuss only informally the implied multiagent transition system and its computations.
A run of the protocol by the miners $\Pi$ results in a sequence of configurations $\rho = c_0, c_1, \ldots$, each encoding the local state of each miner. A miner is \emph{correct} in a run $\rho$ 
if it behaves according to the pseudocode during $\rho$, \emph{faulty} otherwise. As stated above, we assume that there are at most $f < n/3$ faulty miners in any run.
We use $B_p(c)$ to denote the local blocklace of miner $p \in \Pi$ in configuration $c$, $B_p(\rho)$ to denote the blocklace of miner $p$ in the limit, $B_p(\rho) := \bigcup_{c\in \rho} B_p(c)$, and $B(\rho)$ to denote the unions of the blocklaces of all correct miners in the limit, $B(\rho):= \bigcup_{p\in P} B_p(\rho)$, where $P \subseteq \Pi$ is the set of correct miners in run $\rho$.  

We start by showing miner asynchrony (not to be confused with the model of asynchrony), that is, if a miner can create a block, then it can still create it regardless of additional blocks it receives from other miners (Prop. \ref{proposition:miner-asynchrony}). Miner asynchrony combined with the standard notion of \emph{fairness}, that a transition that is enabled infinitely often in a run is eventually taken in the run, implies that once a Cordial Miners block creation transition is enabled then it will eventually be taken (Prop. \ref{proposition:miners-liveness}). We conclude that every miner $p$ correct in a run produces the blocklace of the run, namely $B_p(\rho) = B(\rho)$ (Prop. \ref{proposition:cordial-dissemination}). We can now argue the safety of the Cordial Miners protocols (Prop. \ref{proposition:CM-safety}).

We now proceed to argue the liveness of the Cordial Miners protocols. We show that the Cordial Miners eventual synchrony protocol is leader-live with probability 1 (Prop. \ref{proposition:cordial-leader-liveness-es}).
We note that, following GST, the probability of a leader block being final is at least $\frac{|P|}{n}$, where $P \subseteq \Pi$ is the set of correct miners,
and given that $w=3$, if $\frac{|P|}{n} > \frac{2}{3}$, then the expected latency is at most $3/(2/3)=4.5$ rounds.

The next proposition ensures that all correct miners eventually repel all equivocators and stop observing their blocks. We define an \textbf{equivocator-repelling} block recursively (Def. \ref{definition:equivocator_repelling}), through the set of blocks $B$ that it acknowledges, terminating in an initial block, where $B = \emptyset$.  Note that a block (or blocklace) that is equivocator-repelling may include equivocations, for example, two equivocating blocks each observed by a different block in $B$. However,  once an equivocation by miner $q$ is observed by a block $b$, $q$ would be repelled: Any block that observes $b$ would not acknowledge any $q$-block, preventing any further $q$-blocks from joining the blocklace.  Also note that equivocators are eventually excommunicated since they eventually cannot produce correct blocks (Prop. \ref{proposition:equivocation-free-suffix}).

Lemma \ref{lemma:common-core} claims the existence of a blocklace \textbf{common core}, which is the blocklace-variant of the notion of a common core that appears in~\cite{attiya2004distributed,dolev2016some}. Its proof is an adaptation to the cordial blocklace setting of the common core proof in \cite{dolev2016some}, which in turn is derived from the proof of get-core in \cite{attiya2004distributed}.  Fig. \ref{figure:common-core} illustrates its proof as well as the proof of the following Corollary \ref{corollary:ratified-common-core}, about the existence of a super-ratified common core.

The lemma and corollary require an equivocators-free section of the blocklace, which may be the entire equivocation-free suffix of the blocklace as in the proof.  But the proof also holds if
there is a long enough stretch of rounds without equivocation, in which case a common core also exists.
We conclude that if a Cordial Miners protocol relies on the common core for liveness (Cor. \ref{corollary:liveness_common_core}, Prop. \ref{proposition:cordial-leader-liveness-a}), dissemination, and cordiality are sufficient to ensure it.
Finally, we complete the proof of liveness of the Cordial Miners protocols (Prop. \ref{proposition:CM-liveness}).

This concludes the proof outline that \sys is live and safe and thus completes the proof of Theorem \ref{theorem:CM-safety-liveness}.

\section{Performance Analysis}\label{section:optimizations}
We analyze the performance of \sys assuming the maximum number of Byzantine miners, i.e., $n=3f+1$.
For the good case bit complexity, we assume $f \in O(1)$ and the network is synchronous.

\textbf{Latency} (See Table \ref{table:performance}).
Latency is defined as the number of blocklace rounds between every two consecutive final leaders, i.e., the number of blocklace rounds between two instances where $\tau$ outputs new blocks.
This is also equivalent to the number of communication rounds since we do not use RB to disseminate blocks. The good case latency for both models is simply the wavelength.

For the expected case, in the asynchronous instance of the protocol, each wave $w$ consists of $5$ rounds. According to Lemma \ref{lemma:common-core}, the probability that the leader block is final in the first round $r$ of $w$, namely that a supermajority of the blocks in $r+4$ each super-ratify the leader block at $r$ is $\frac{2}{3}$.
Therefore, in the expected case a leader block is final every $1.5$ waves, and therefore the expected latency is $1.5 w = 7.5$ rounds of communication.

The adversary can equivocate or not be cordial up to $f$ times, but after each Byzantine process $p$ equivocates, all correct processes eventually detect the equivocation and do not consider $p$'s blocks as part of their cordial rounds when building the blocklace.
Thus, in an infinite run, equivocations do not affect the overall expected latency.

In the ES version, each wave $w$ consists of $3$ rounds.
The probability that the leader block is final is if the leader block is created by a correct miner, i.e., the probability is $\frac{2}{3}$, i.e., same as asynchrony.
Thus, in the expected case, the latency is $1.5w = 4.5$ rounds.

\noindent\textbf{Bit complexity.} 
An equivocator is eventually excommunicated, and therefore eventually the number of equivocating blocks that are disseminated is limited (see Prop. \ref{proposition:equivocation-free-suffix}). Each block in the blocklace is linear in size since it has a linear number of hash pointers to previous blocks.
A Byzantine miner can cause the block it creates to be sent to all miners by all the other correct miners, causing the block's bit complexity to be $O(n^3)$ per such block.
Thus, in the worst case, where $f \in \Theta(n)$, the asymptotic bit complexity is $O(n^3)$ per block.
But, since the block size is $O(n)$, we can batch $O(n)$ transactions in it without increasing its asymptotic size. 
Therefore, we can amortize the bit complexity by a linear factor for each transaction, causing the amortized bit complexity per transaction to be $O(n^2)$ in the worst case.

For the good case in the ES version, where $f \in O(1)$ and the network is synchronous after GST, every block created by a correct miner is sent once from its creator to the other miners.
Miners wait for $\textit{timeout}$ time after a round $r$ is cordial before they move to the next round, which ensures that all blocks sent by correct miners in round $r$ arrive to all other correct miners before they move to round $r+1$.
Therefore, blocks by correct miners in round $r+1$ observe all blocks by correct miners in round $r$.
Thus, the Byzantine miners can cause only a constant number of blocks per round to be sent by every correct miner to every other correct miner.
Therefore, the bit complexity of sending each block in the good case is $O(n^2)$, and by batching $O(n)$ transaction per block, we get an amortized bit complexity of $O(n)$ per transaction.

\section{Related Work} \label{sec:related-work}

The use of a DAG-like structure to solve consensus has been introduced in previous works, especially in asynchronous networks.
Hashgraph~\cite{RN284} builds an unstructured DAG, with each block containing two references to previous blocks, and on top of the DAG, the miners run an inefficient binary agreement protocol.
This leads to expected exponential time complexity.
Aleph~~\cite{gkagol2018aleph} builds a structured round-based DAG, where miners proceed to the next round once they receive $2f+1$ DAG vertices from other miners in the same round.
On top of the DAG construction protocol, a binary agreement protocol decides on the order of vertices to commit.
Blockmania~\cite{danezis2018blockmania} uses a variant of PBFT~\cite{RN581} in the ES model and also uses reliable broadcast to disseminate blocks.
Both protocols have higher latency than \sys since they use RB.
GHOST~\cite{sompolinsky2015secure}, IOTA~\cite{RN349}, and Avalanche~\cite{rocket2019scalable} are DAG protocols for the permissionless model.

As mentioned in the introduction, the two state-of-the-art DAG-based protocols are DAG-Rider~\cite{keidar2021need} and Bullshark~\cite{giridharan2022bullshark}.
DAG-Rider is a BAB protocol for the asynchronous model in which the miners jointly build a DAG of blocks, with blocks as vertices and pointers to previously created blocks as edges, divided into strong and weak edges.
Strong edges are used for the commit rule, and weak edges are used to ensure fairness.
Narwhal~\cite{danezis2021narwhal} is an implementation based on DAG-Rider for a relaxed networking model and works well assuming messages arrival is not bounded, but also not controlled by the adversary.
Tusk~\cite{danezis2021narwhal} is a similar consensus protocol to DAG-Rider built on top of Narwhal.
Bullshark~\cite{giridharan2022bullshark} is a variation of DAG-Rider designed for the ES model with about half the latency of DAG-Rider.
\sys outperform these protocols in terms of latency (for the same message complexity).
Other DAG-based protocols include~\cite{chockler1998adaptive,dolev1993early}, which are for a non-Byzantine failure model.

Another category of Byzantine consensus protocols is Leader-based. Examples include PBFT~\cite{RN581}, Tendermint~\cite{buchman2016tendermint}, HotStuff~\cite{yin2019hotstuff,malkhi2023hotstuff}, and VABA~\cite{abraham2019asymptotically}.
In these protocols, a designated leader proposes a block, sends them to the miners, and collects votes on its proposal, and a Byzantine leader can result in wasted time in which no blocks are output.
Another difference is that these protocols are unbalanced in terms of the network as the leader is in charge of disseminating its block, collecting votes, and disseminating them, while the other miners only need to vote.
On the other hand, DAG-based protocols like \sys are symmetric in that all miners perform exactly the same tasks.

\section{Conclusion}
\label{sec:futureDirections}
We presented \sys, a family of low-latency, high-efficiency consensus protocols with instances for eventual synchrony and asynchrony.  \sys achieve that by forgoing Reliable Broadcast and using the blocklace for the three major tasks of consensus -- dissemination, equivocation exclusion, and ordering.

\bibliography{bib}

\appendix
\section{Formal Model} \label{appendix-section:model}

The following is a mathematical formal definition of the cordial miners consensus protocols.

\begin{definition}[Block, Acknowledge]\label{definition:block}
    A \temph{block} $b$ is a triple $b=(p,a,H)$ signed by $p$, referred to as a \temph{$p$-block}, s.t. $p \in \Pi$ is the miner that creates $b$, $a \in \calA$ is the \temph{payload} of $b$, and $H$ is a finite set of hash pointers to blocks.
    Namely, for each $h \in H$, $h=\textit{hash}(b')$ for some block $b'$.
    In which case we also say that $b$ \temph{acknowledges} $b'$.
    If $H = \emptyset$ then $b$ is \temph{initial}. 
\end{definition}

\begin{definition}[Blocklace]\label{definition:blocklace}
    Let $\calB$ be the maximal set of blocks over $\Pi$, $\calA$, and $\textit{hash}$ for which the induced directed graph $(\calB,\calE)$ is acyclic. A \temph{blocklace} over $\calA$ is a set of blocks $B \subseteq \calB$.
\end{definition}

\begin{definition}[$\succ$, Observe]\label{definition:observe}
    Given two blocks $b,b'$, the strict partial order $\succ$  is defined by $b'\succ b$ if there is a nonempty path from $b'$ to $b$.  A block  $b'$ \temph{observes} $b$ if $b'\succeq b$. Given a blocklace $B$, Miner $p$ \temph{observes $b$ in} $B$ if there is a $p$-block $b' \in B$ that observes $b$.
    A group of miners $Q \subseteq \Pi$ \temph{observes $b$ in} $B$ if every miner $p \in Q$ observes $b$.
\end{definition}

\begin{definition}[Equivocation, Equivocator]\label{definition:equivocation}
    A pair of $p$-blocks $b\ne b'\in \calB$, $p \in \Pi$, form an \temph{equivocation} by $p$ if they are not consistent wrt $\succ$, namely $b' \not\succ b$ and $b \not\succ b'$. A miner $p$ is an \temph{equivocator in} $B$, $\textit{equivocator}(p,B)$, if $B$ has an equivocation by $p$. 
\end{definition}

\begin{definition}[Approval]\label{definition:approval}
    Given blocks $b, b'\in \calB$, the block $b$ \temph{approves} $b'$ if $b$ observes $b'$ and does not observe any block $b''$ that together with $b'$ forms an equivocation. A miner $p \in \Pi$ \temph{approves $b'$ in $B$} if there is a $p$-block $b\in B$ that approves $b'$. A set of miners $Q \subseteq \Pi$ \temph{approve $b'$ in $B$} if every miner $p \in Q$ approves $b'$ in $B$.
\end{definition}

\begin{definition}[Closure, Closed, Tip]\label{definition:closure}
    The \temph{closure of $b \in \calB$ wrt $\succ$} is the set $[b] := \{b'\in \calB : b \succeq b' \}$. The \temph{closure of $B\subset \calB$  wrt $\succ$} is the set $[B] := \bigcup_{b \in B} [b]$.
    A blocklace $B \subseteq \calB$ is \temph{closed} if $B = [B]$. 
    A block $b \in \calB$ is a \temph{tip} of $B$ if $b \notin [B \setminus \{b\}]$.
\end{definition}

\begin{definition}[Block Depth/Round, Blocklace Prefix \& Suffix] \label{definition:depth}
    The \temph{depth} (or \temph{round}) of a block $b \in \calB$, $\textit{depth}(b)$, is the maximal length of any path of pointers emanating from $b$.  For a blocklace $B\subseteq \calB$ and $d\ge 0$, the \temph{depth-$d$ prefix of} $B$ is $B(d):= \{b \in B : \textit{depth}(b) \le d \}$, and the
    \temph{depth-$d$ suffix of} $B$ is  $\bar{B}(d):= B  \setminus  B(d)$.
\end{definition}

\begin{definition}[Supermajority] \label{definition:supermajority}
    A set of miners $P \subset \Pi$ is a \temph{supermajority} if $|P| > \frac{n+f}{2}$.  A set of
    blocks $B$ is a \temph{supermajority} if the set of miners $P=\{p\in \Pi : \exists b \in B \text{ is a $p$-block}\}$ is a supermajority.
\end{definition}

\begin{definition}[Ratified and Super-Ratified Block]\label{definition:ratified-block}
    A block $b \in \calB$  is (\ia) \temph{ratified by a set of blocks} $B \subseteq \calB$, 
    if $[B]$ includes a supermajority of blocks that approve $b$; (\ib)   \temph{ratified by a block} $b$ if it is ratified by the set of blocks $[b]$; and (\ic) \temph{super-ratified} by blocklace $B \subset \calB$ if $[B]$ includes a supermajority of blocks, each of which ratifies $b$ 
\end{definition}

\begin{definition}[Wavelength, Leader Selection Function, Leader Block] \label{def:leaderFunc}
	Given a \temph{wavelength} $w \ge 1$, a \temph{leader selection function} is a partial function $l: \calN \mapsto \Pi$ satisfying (\ia) \textbf{coverage:} $\forall r \in \calN \colon l(r) \in \Pi$ if $r \temph{ mod } w = 0$ else $l(r) = \bot$ and (\ib) \textbf{fairness:} with probability 1 $\forall r \in \mathbb{N}, p \in \Pi~ \exists r'>r: l(r') = p$.
	A $p$-block $b$ is a \temph{leader block} if $l(\textit{depth}(b)) = p$.
\end{definition}

\begin{definition}[Final Leader Block]\label{definition:final-leaders}
    Let $B \subseteq \calB$ be a  blocklace. A leader block $b \in B$ of round $r$  is \temph{final} in $B$ if it is super-ratified in $B(r+w-1)$. 
\end{definition}

\begin{definition}[Cordial Block, Blocklace]\label{definition:cordial}
	A block $b \in \calB$ of round $r$ is \temph{cordial} if $r=1$ or it acknowledges blocks by a supermajority of miners of round $r-1$.  A blocklace $B\subset \calB$ is \temph{cordial} if all its blocks are cordial.
\end{definition}

\begin{definition} \label{definition:correct-block}
    A $p$-block $b \in \calB$ in \temph{correct}, if it is cordial and $p$ does doe equivocate in $[b]$.
\end{definition}

\begin{definition}[Disseminating]\label{definition:disseminating}
	Given a blocklace $B\subseteq \calB$, a set of miners $P \subseteq \Pi$ is \temph{mutually disseminating} in $B$, or \temph{disseminating} for short, if for any $p,q \in P$ and any $p$-block $b\in B$ there is a $q$-block $b' \in B$ such that $b' \succ b$. The blocklace $B$ is \temph{disseminating} if it has a disseminating supermajority.
\end{definition}

\begin{definition}[Shared Random Coin]\label{definition:shared-random-coin}
	Assume some $d>0$ and let $S = \{\textit{toss\_coin}(p_s,d): p \in P\}$ for a set of miners $P\subseteq \Pi$, $|P| > f+1$. 
	For the shared random coin, the function   $\textit{combine\_tosses}$ has the following properties:
	\begin{description}
		\item[Agreement] If both $S', S'' \subseteq S$ and both $|S'|, |S''| > f+1$, then\\
		$\textit{combine\_tosses}(S',d) = \textit{combine\_tosses}(S'',d)$
		\item [Termination] 
		$\textit{combine\_tosses}(S,d) \in \Pi$.
		\item[Fairness] The coin is fair, i.e., for every set $S$ computed as above and any $p \in \Pi$, the probability that $p=\textit{combine\_tosses}(S,d)$ is  $\frac{1}{n}$.
		\item[Unpredictability]  If $S' \subset S$, $|S'|< f+1$, then
		the probability that the adversary can use $S'$ to guess the value of  $\textit{combine\_tosses}(S,d)$ is less than $\frac{1}{n} + \epsilon$.
	\end{description}
\end{definition}

\begin{definition}[Equivocator-Repelling]\label{definition:equivocator_repelling}
	Let  $b\in \calB$ be a $p$-block, $p\in \Pi$, that acknowledges a set of blocks $B \subset \calB$.
	Then $b$ is \temph{equivocator-repelling} if $p$ does not equivocate in $[b]$ and
	all blocks in $B$ are equivocator-repelling.
	A blocklace $B$ is \temph{equivocator-repelling} if every block $b \in B$ is equivocator-repelling.
\end{definition}

\clearpage
\section{Figures}\label{appendix-section:figures}
\begin{figure}[ht]
	\centering
	\includegraphics[width=0.7\textwidth]{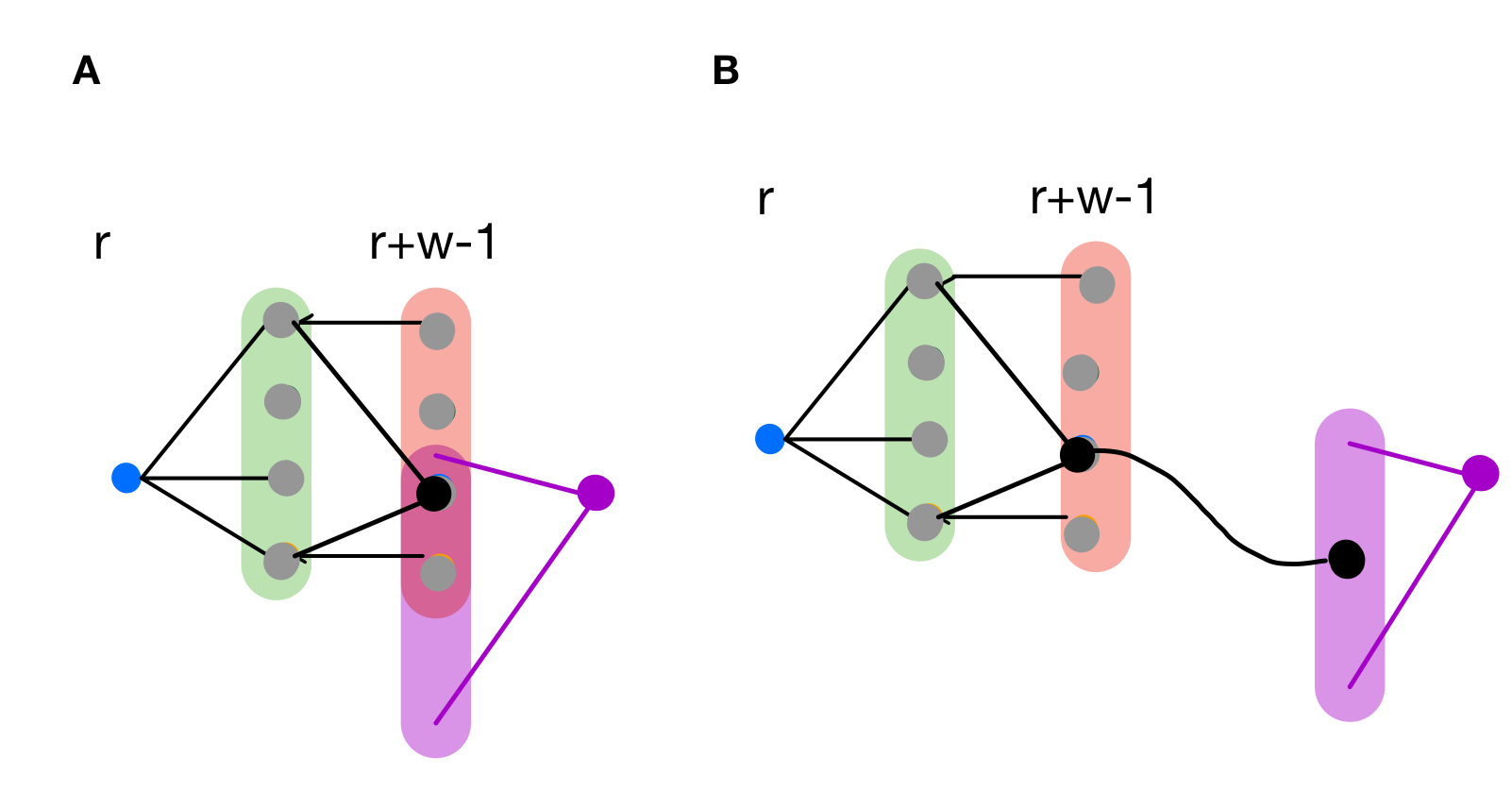}
	\caption{Finality of a Super-Ratified Leader (Definition \ref{definition:final-leaders}):  Assume
		that a leader block (blue dot) is super-ratified. A ratifying supermajority is represented by a thick red line, each member of which observes a possibly different approving supermajority represented by a green thick line.  We show that the blue leader is ratified by any subsequent cordial leader.
		(A) The successive cordial leader (purple dot) is one round following the ratifying supermajority. Being cordial, it observes a supermajority (thick purple line)  that must have an intersection (black dot) with the ratifying supermajority, hence it observes an approving supermajority and thus ratifies the blue leader.
		(B) A successive leader is more than one round following the ratifying supermajority.
		Being cordial, it observes a supermajority (thick purple line).  There must be a correct miner common to the purple and red supermajority, with blocks in both (black dots); being a correct miner, its later block observes the earlier block (black line).  Hence the purple leader observes the approving supermajority (via black lines) and hence ratifies the blue leader.
	}
	\label{figure:finality}
\end{figure}

\begin{figure}[t]
	\centering
	\includegraphics[width=0.3\textwidth]{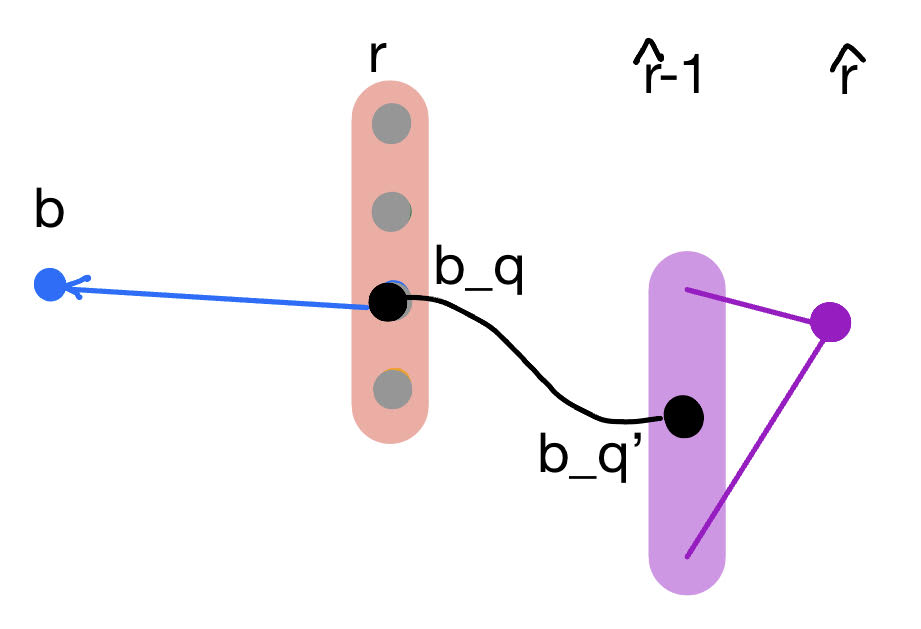}
	\caption{Liveness Condition, Proposition~\ref{proposition:liveness-condition}.}
	\label{figure:liveness-condition}
\end{figure}

\begin{figure*}[t]
	\centering
	\includegraphics[width=0.8\textwidth]{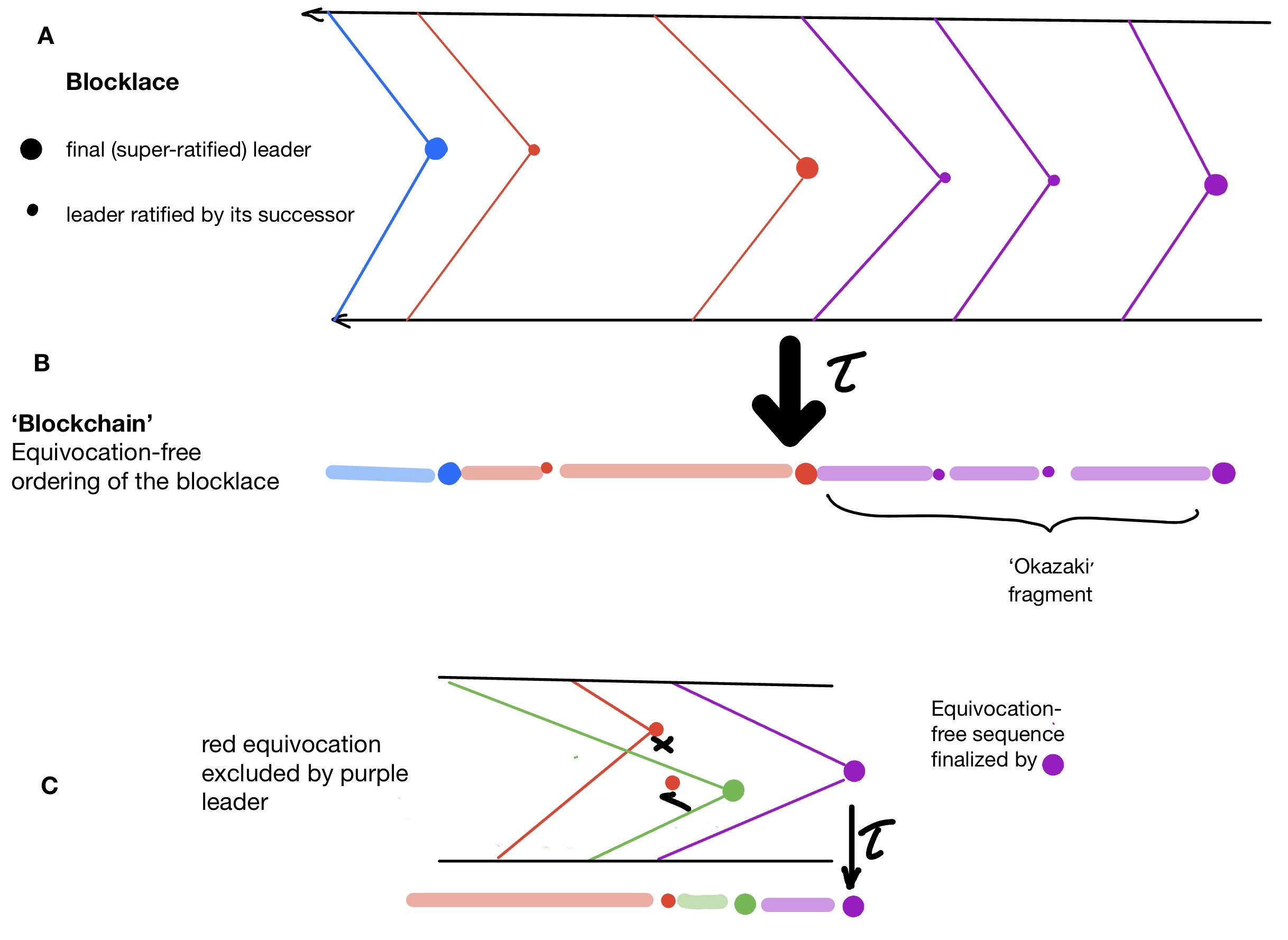}
	\caption{ \textbf{The Operation of $\tau$, Safety and Liveness:}
		(A) \emph{The Input of $\tau$}: A blocklace with final leaders (large dots) and leaders ratified by their successors (small dots). Each leader observes the portion of the blocklace below it (including the lines emanating from it).
		(B) \emph{The Output of $\tau$}: A sequence of blocks consisting of  fragments. 
		The sequence of fragments is computed recursively backward, starting from the last final leader, and back from each leader to the previous leader it ratifies. The input to computing the fragment consists of the portion of the blocklace observed by the current leader but not observed by the previous ratified leader. The output from each fragment is a sequence of blocks computed forward by topological sort of the input blocklace fragment,  respecting $\succ$ and using the leader of the fragment to resolve and exclude equivocations. Final leaders are final, hence the backward computation starting from the last purple final leader need not proceed beyond the recursive call to the previous red final leader, as the output sequence up to the previous final leader has already been computed by the previous invocation of $\tau$. \textbf{Safety Requirement:} A final leader (large dot) is ratified by any subsequent leader (large or small dot). 
		\textbf{Liveness Requirement:} Any leader will eventually have a subsequent final leader (large dot) with probability 1.
		(C) \emph{Leader-Based Equivocation Exclusion}: The green fragment created by the green leader includes the $V$-marked red block, since the green leader does not observe the red equivocation.  However, the red $X$-marked red block is excluded from the purple fragment created by the purple leader, since the purple leader observes the equivocation among the two red blocks.}
	
	\label{figure:tau}

\end{figure*}

\begin{figure*}[ht]
	\centering
	\includegraphics[width=0.6\textwidth]{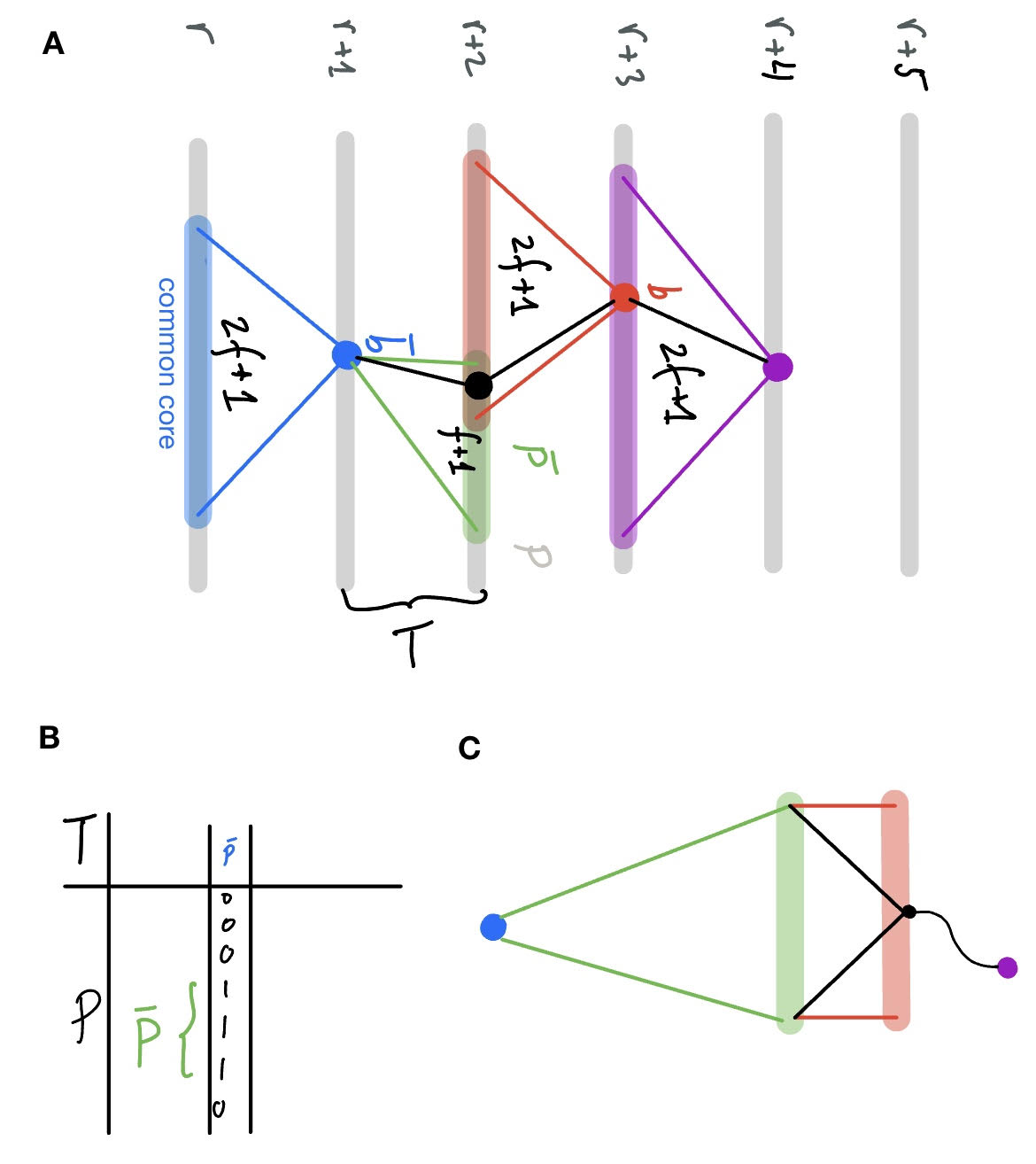}
	\caption{
		\textbf{Common Core, Ratified Common Core, Safety and Liveness of Decision Rule for Asynchrony:}
		(A) \emph{Proof of Lemma \ref{lemma:common-core} and Corollary \ref{corollary:ratified-common-core}}: Rounds $r$ to $r+3$ relate to the proof of Lemma \ref{lemma:common-core}, where the existence of a common core at round $r$ is established. Round $r+4$ relates to Corollary \ref{corollary:ratified-common-core}, which establishes that all cordial blocks at round $r+4$ ratify all members of the common core of round $r$ via a supermajority at round $r+3$. 
		(B) \emph{The common-core table $T$} used in the proof to relate rounds $r+1$ and $r+2$. 
		(C) \emph{The decision rule for asynchrony}: Protocol wavelength is 5. \textbf{Liveness:} Common-core ensures that the blue leader at round $r$ is super-ratified by a red supermajority at round $r+4$ with probability $\frac{2f+1}{3f+1}$, thus ensuring liveness and expected latency of 7.5 rounds.  \textbf{Safety:} A blue leader is approved by every cordial block at round $r+3$ (green) and hence is ratified by every cordial block at round $r+4$ (red) and beyond.}
	\label{figure:common-core}
	
\end{figure*}

\clearpage

\section{Proofs} \label{appendix-section:proofs}

In this section, we provide the full proofs deferred from the paper.

\CordialBlocklaceLeaderSafety*
\begin{proof}[Proof of Proposition \ref{proposition:cordial-safety}]
	Let $B$ be a cordial blocklace and $b \in B$ (blue dot in Fig. \ref{figure:finality}) a final leader block of round $r$ in $B$. We have to show $b$ is ratified by any subsequent leader block in $B$.
	We consider Definition \ref{definition:final-leaders}, and in reference to the two cases in Fig. \ref{figure:finality}.
	Consider a leader block  $b$ of round $r$ (black dot) and a leader block $b'$ (purple dot) of a round $r' > r+w$.  There are two cases:
	\begin{enumerate}
		\item[A.]  There is an overlap between the supermajority observed by $b'$ at round $r'-1$ (thick purple line) and the supermajority that ratifies $b$ (thick red line).  Then there is a block $b''$ (black dot) shared by both, hence $b'$ observes $b''$, which observes a supermajority (thick green line) that approves $b$, hence $b'$ ratifies $b$.
		\item[B.] The supermajority observed by $b'$ at round $r'-1$ (thick purple line) is of a later round than the members of the supermajority that ratifies $b$ (thick red line).  By counting, there is a non-equivocating miner $p \in Pi$ with a block $b_1$ in the purple supermajority and a block $b_2$ in the red supermajority (black dots).  Since $p$ is non-equivocating, $b_1$ observes $b_2$.  Hence, $b'$ observes $b_1$, which observes $b_2$, which observes the green supermajority that approves $b$, hence $b'$ ratifies $b$.
	\end{enumerate}
	This completes the proof.
\end{proof}

\BlocklaceLeaderLivenessCondition*
\begin{proof}[Proof of Proposition \ref{proposition:liveness-condition}]
	Let $B\subset \calB$ be a cordial blocklace and $P \subseteq \Pi$ a  supermajority of miners non-equivocating and disseminating in $B$.  Let $b$ be a $p$-block by a miner $p \in P$ (blue dot in Fig. \ref{figure:liveness-condition}).  As $P$ are disseminating in $B$, then for every $q \in P$ there is a first $q$-block $b_q \in P$ that observes $b$; let $r$ be the maximal round of any of these blocks (thick horizontal red line).
	By assumption, $B$ has a final leader block $\hat{b}$ of round $\hat{r}>r$ (purple dot).  As $\hat{b}$ is cordial, it must observe a block $b'_q$ of depth $\hat{r}-1$ of a miner $q \in P$ (black dot).  As $q$ is non-equivocating, there is a (possibly empty) path from  $b'_q$ to $b_q$ (black path among black dots), and from there to $b$ (blue line).  Hence $\hat{b}$ observes $b$.
\end{proof}

\begin{observation}[Approving an Equivocation] \label{observation:approve-da}
	If miner $p \in \Pi$ approves an equivocation $b_1, b_2$ in a blocklace $B \subseteq \calB$, then $p$ is an equivocator in $B$.
\end{observation}
\begin{proof}[Proof of Observation~\ref{observation:approve-da}]
By way of contradiction, assume that $p$ approves an equivocation $b_1, b_2 \in B$ via two blocks, $b_3$ and $b_4$ in $B$, respectively, that do not constitute an equivocation.  So w.l.o.g.\  assume that $b_3$ observes $b_4$.
However, since $b_4$ observes $b_2$, then $b_3$ also observes $b_2$, and hence does not approve $b_1$.  A contradiction.
\end{proof}
\begin{lemma}[No Supermajority Approval for Equivocation]\label{lemma:no-double-majority}
    For every two equivocating blocks in a blocklace, at most one can have a supermajority approval.
\end{lemma}
\begin{proof}[Proof of Lemma~\ref{lemma:no-double-majority}]
Assume that $b_1, b_2$ are equivocating blocks, and that there are two supermajorities  $Q_1, Q_2 \subseteq  P$, where $Q_1$ approves $b_1$ and
$Q_2$ approves $b_2$.
A counting argument shows that two supermajorities must
have in common at least one correct miner. Let $p \in Q_1 \cap Q_2$
be such a correct miner.  Since $p\in Q_1$ it approves $b_1$ and since $p\in Q_2$ it approves $b_2$ by construction. Observation \ref{observation:approve-da} shows that an miner that approves an equivocation must be a
an equivocator, hence $p$ is an equivocator.  A contradiction.
\end{proof}

\begin{observation}[Dissemination is Unbounded]\label{observation:infinite-dissemination}
	If a set of miners $P \subseteq \Pi$, $|P|>1$, are  disseminating in a blocklace $B$ that includes a $p$-block, $p \in P$, then $B$ is infinite and  any suffix of $B$ has blocks by every member of $P$.
\end{observation}
\begin{proof}
Under the assumptions of the observation, every correct miner $q$ will receive $b$ and create $q$-block observing $b$.  $B$ cannot be finite since any `last' block by a correct miner must be followed by blocks by all correct miners, so it can't be last.  And for the same reason any suffix has blocks by all correct miners.
\end{proof}

\MonotonicityofTau*
\begin{proof}[Proof of Proposition \ref{proposition:tau-finality}]
	Let $B, B_1, B_2$ be blocklaces as assumed by the Proposition.  If $B_2$ has no final leader then $\tau(B_2)$ is the empty sequence and the proposition holds vacuously.
	Let $\hat{b}_2$ be the last final leader of $B_2$ and $\hat{b}_1$ be the last final leader of $B_1$.
	Note that according to Definition\ref{definition:tau},
	$\tau(B_1)$ calls $\tau'(\hat{b}_1)$ and $\tau(B_2)$ calls $\tau'(\hat{b}_2)$.  Let $b_1,b_2\ldots,b_k$, $k \ge 2$, be the sequence of ratified leaders in the recursive calls of the execution of  $\tau'(\hat{b}_1)$, starting with $b_1 = \hat{b}_1$.  We argue that  $\hat{b}_2$ is called in this execution, namely  $\hat{b}_2 = b_j$ for some $j \in [k]$.  
	Note that according to Proposition \ref{proposition:cordial-safety}, $\hat{b}_1=b_1$, being cordial, ratifies  $\hat{b}_2$.  Let $j \in [k]$ be the last index for which $b_j$ ratifies  $\hat{b}_2$.
	We argue by way of contradiction that $b_{j+1}=\hat{b}_2$. Consider three cases regarding the relative depths of $b_{j+1}$ and $\hat{b}_2$:
	\begin{itemize}
		\item[=] Note that two different blocks of the same depth cannot observe each other: If only one observes the other, it is one deeper than the other; if both observe each other they form a cycle, which is impossible.   Since both $b_{j+1}$ and $\hat{b}_2$ are leader blocks of the same depth, then they must be by the same leader, and hence, being different blocks by the same miner that do not observe each other, they form an equivocation. By assumption,  both are ratified, implying that both have supermajority approval, contradicting the assumption that there is a supermajority of correct miners in $B$. 
		\item[>] If $b_{j+1}$ is deeper than $\hat{b}_2$,
		then by Definition \ref{definition:tau}, $\tau'$ elects the first leader ratified by the current leader $b_j$, and hence cannot prefer calling $b_{j+1}$  over $\hat{b}_2$, which precedes it by assumption.
		\item[<] If $\hat{b}_2$ is deeper, then Proposition \ref{proposition:cordial-safety} implies that $b_{j+1}$ ratifies $\hat{b}_2$, in contradiction to the assumption that  $b_j$ is the last leader in the list that ratifies $\hat{b}_2$.
	\end{itemize}
	Hence $\hat{b}_2$ is included in the recursive calls of $\tau'(\hat{b}_1)$, which, according to Definition \ref{definition:tau} of $\tau$, implies that
	$\tau(B_2) \preceq \tau(B_1)$.
\end{proof}

\begin{observation}[Consistent triplet]\label{observation:consistent-triplet}
	Given three sequences $x, x', x''$, if both $x' \preceq x$ and $x'' \preceq x$ then $x'$ and $x''$ are consistent.
\end{observation}
\begin{proof} [Proof of Observation~\ref{observation:consistent-triplet}]
 Let $|x'|=k'$  and $|x''|=k''$.  By assumption, $x'$ consists of the first $k'$ elements of $x$ and
 $x''$ consists of the first $k''$ elements of $x$. Wlog assume $k'\le k''$.  Then $x'$ consists of the first $k'$ elements of $x''$, and hence $x'\preceq x''$.
\end{proof}

\TauSafety*
\begin{proof}[Proof of Proposition \ref{proposition:tau-safety}]
	By monotonicity of $\tau$ (Prop.  \ref{proposition:tau-finality}), both $\tau(B_1) \preceq \tau(B_1 \cup B_2)$ and $\tau(B_2) \ \preceq \tau(B_1 \cup B_2)$. By Observation \ref{observation:consistent-triplet}, $\tau(B_1)$ and $\tau(B_2)$ are consistent.
\end{proof}

\TauOutput*
\begin{proof}[Proof of Observation \ref{observation:tau-output}]
	Since $b$ is observed by a final leader in $B$, it is also observed by the last final leader of $B$.
	Consider the recursive construction of $\tau(B)$.  If in its last recursive call $\tau'(b')$, $b'$ observes $b$, then  by definition of $\tau$, $b \in \tau(b')$ and hence $b \in \tau(B)$.  Otherwise,
	consider the first recursive call $\tau'(b')$ by $\tau'(b'')$ in which $b''$ observes $b$ but $b'$ does not observe $b$. Then  by definition of $\tau'$, $b \in \tau'(b'')$ and hence $b \in \tau(B)$. 
\end{proof}

\TauLiveness*
\begin{proof}[Proof of Proposition \ref{proposition:tau-liveness}]
	Let $B$ be as assumed and $b \in B$.  As $B$ is leader-live, there is a final leader block $b'$ that observes $b$.  Let $i\ge 1$ be an index for which $B_i$ includes $b'$.
	Consider the call $\tau(B_i)$.  Since $b$ is a final leader in $B_i$, then according to Observation $\ref{observation:tau-output}$, the output of $\tau(B_i)$ includes $b$.
\end{proof}

\SufficientCondition*
\begin{proof}[Proof of Theorem \ref{theorem:tau-conditions-safety-liveness}]
	Let $P \subseteq \Pi$ be the correct miners in a run of the protocol that produce in the limit the blocklace $B$. The protocol is safe since the local blocklaces of any two miners in $p, q \in P$ at any time are subsets of $B$, hence by Proposition \ref{proposition:tau-safety} the outputs of $p$ and $q$ are consistent. The protocol is live since by Proposition \ref{proposition:tau-liveness}, every block  $b\in B$ will be output by every  correct miner $p \in P$.
\end{proof}

\safetyAndLiveness*
\begin{proof}[Proof Outline of Theorem~\ref{theorem:CM-safety-liveness}]
	We prove two propositions that together establish the Theorem: Proposition \ref{proposition:CM-safety} shows that the two Cordial Miners protocols are safe and  Proposition \ref{proposition:CM-liveness} shows that they are live.
\end{proof}

\begin{proposition}[Miner Asynchrony]\label{proposition:miner-asynchrony} 
	If a miner can create a block (Line \ref{alg:createBlock}) then it can create it also after receiving blocks from other miners.
\end{proposition}

\begin{proof}[Proof of Proposition \ref{proposition:miner-asynchrony}]
	Examination of the \textit{completed\_round} procedures of Alg. \ref{alg:protocols}, which gate block creation in Alg. \ref{alg:dissemination}, Line \ref{alg:createBlock}, shows that if it holds for a blocklace it holds after blocks by other miners are received and buffered or added to the local blocklace.
\end{proof}

\begin{proposition}[Miners Liveness]\label{proposition:miners-liveness}
	In a fair run of a Cordial Miners protocol with correct miners $P \subseteq \Pi$, 
	if there is a configuration for which $\textit{completed\_round}()\ge d$ (Line \ref{alg:completed-round})  for $d\ge 0$ and for the blocklace of every miner $p\in P$, then there is a subsequent configuration for which  $\textit{completed\_round}()\ge d+1$ for the blocklace of every miner $p\in P$.
\end{proposition}

\begin{proof}[Proof of Proposition \ref{proposition:miners-liveness}]
	We show by induction on the round number.  Consider a configuration $c$ in which the depth of the last completed round in the blocklace of all miners be $d \ge 0$.
	If $d = 0$ then 
	the \emph{completed\_round}$()$ call (Line \ref{alg:completed-round}) returns $0$ and $p$ can create an initial block (Line \ref{alg:createBlock}) with no predecessors (Line \ref{alg:tips}). 
	Assume $d >0$.  Consider a miner $p \in P$ that has not yet created a block of depth $d+1$ in $c$.
	Then the condition $\emph{completed\_round}() \ge r$  (Line \ref{alg:completed-round}) holds for $r =d$, and the transition to create the next block is enabled.
	By miner asynchrony (Proposition \ref{proposition:miner-asynchrony}), such a transition is enabled indefinitely, and by the fairness assumption, it is eventually taken, in which $p$ sends a new $p$-block $b$ of depth $d+1$ to all other miners (Line \ref{alg:send-block}), and includes, for each miner $q$, all the blocks in the closure of $b$ that $q$ might not know of, based on the communication history of $p$ with $q$.  By assumption, all said messages among correct miners eventually arrive at their destination.
	Hence there is some subsequent configuration $c'$ in which every correct miner receives a $d+1$-depth $q$-block $b$ from every other correct miner $q$, as well as all preceding blocks to $b$.
	Hence in $c'$, $\emph{completed\_round}() \ge r$ holds for $r =d+1$ for every correct miner.
\end{proof}

\begin{proposition}[Cordial Miners Dissemination]\label{proposition:cordial-dissemination}
	In a run $\rho$ of a Cordial Miners protocol with correct miners $P \subseteq \Pi$, $B(\rho) = B_p(\rho)$ for every  $p \in P$.
\end{proposition}

\begin{proof}[Proof of Proposition \ref{proposition:cordial-dissemination}]
	Given a run $\rho$ of a Cordial Miners protocol with correct miners $P \subset \Pi$, we 
	have to show that for any $p, q \in P$,  a configuration $c$ of the run, and a block $b$  in the blocklace of $p$ is configuration $c$, there is a subsequent configuration $c'$ of the run in which $b$ is in the local blocklace of $q$. 
	By miners liveness (Proposition \ref{proposition:miners-liveness}), for any miner $q$, there is a subsequent configuration by which miner $p$ sends a block to miner $q$.  According to the Cordial Dissemination clause (Line \ref{alg:send-block}),  when $p$ sends to $q$ a $p$-block of round $\ge \textit{depth}(b)+1$ $p$-block, it will also send to $q$ all
    blocks that $b$ depends upon, minus any blocks known to $q$ according to the most recent $q$-block received by $p$.  Hence there is a subsequent configuration to $c'$ in which $b$ is included in the blocklace of $q$.
\end{proof}

\begin{proposition}[Cordial Miners Protocol Safety]\label{proposition:CM-safety} 
	The Cordial Miners protocols for asynchrony and eventual synchrony are safe.
\end{proposition}

\begin{proof}[Proof of Proposition \ref{proposition:CM-safety}]
	According to Proposition \ref{proposition:cordial-dissemination}, in any computation of a Cordial Miners protocol, the local blocklace of any two correct miners $p,q \in \Pi$ is the same blocklace $B$.
	Hence, in any configuration of the computation, the local blocklaces of $p$ and $q$ are subsets of $B$, and hence according to Proposition \ref{proposition:tau-safety}, their outputs at that configuration are consistent, which is the safety requirement of ordering consensus protocols (Def. \ref{definition:safety-liveness}).  Hence the Cordial Miners protocols are safe.
\end{proof}

\begin{proposition}[Leader-Liveness of Cordial  Miners Eventual Synchrony Protocol]\label{proposition:cordial-leader-liveness-es}
	The blocklace produced by a run of a Cordial Miners eventual synchrony protocol is leader-live with probability 1, if $\textit{timeout} > \Delta$.
\end{proposition}

\begin{proof}[Proof of Proposition \ref{proposition:cordial-leader-liveness-es}]
Let $B$ be the cordial blocklace produced by a run of a Cordial Miners eventual synchrony protocol, $P\subseteq \Pi$ the supermajority of miners correct in the run, and let $r>0$ be any round for which the $r-1$ suffix of $B$, $\bar{B}(r-1)$, is equivocation-free, $r \text{ mod } w = 0$, where $w= 3$ (Line \ref{alg:w-2}).

Let $r'>r$ be any round following network synchronization (GST) where $r \bmod w =0$, and assume the honest miners are in $r'$.
Since leader selection is pseudorandom (Line \ref{alg:leader_es}), and $P$ is a supermajority, there is a probability of $\frac{|P|}{n}$ that the $r'$-leader $q$ is correct, namely $q \in P$.
Let $t$ be the first time in which the blocklace of round $r'$ at some honest miner $p$ is cordial. 
Honest miners wait until the condition for block finality is achieved before proceeding to the next round, or for a \textit{timeout}, which is the estimation of the network delay $\Delta$ (Line~\ref{alg:timeoutES}). By the model, we assume this estimation is accurate after GST.
Since the blocks sent by honest miners arrive within $\Delta$, by $t+\Delta$, $p$ receives the blocks of all honest miners in $r'$, including $q$.
Then, for every $p \in P$, the $r'$-depth $q$-block $b$ is approved by the $r'+1$ $p$-block and super-ratified by the $r'+2$-round block of $p$.
Hence, the blocks of the correct miners $P$ satisfy the conditions of $b$ being a final leader.
As this holds also for the leader of any round following $r$, the probability that for any depth $r'\ge r$, a leader in $\bar{B}(r')$ has a final leader is 1, hence $B$ is leader-live with probability 1.
Hence the condition of Proposition \ref{proposition:liveness-condition},  that for every $r>0$, $B$ has a final leader block of some round $r'>r$ with probability 1, is satisfied and we conclude that $B$ is leader-live  with probability 1.
\end{proof}

\begin{proposition}[Equivocators-Free Suffix]\label{proposition:equivocation-free-suffix}
	Let $B$ be an equivocator-repelling and cordial blocklace and $P \subset \Pi$ the set of miners that do not equivocate in $B$ and are disseminating in $B$. If $P$ is a supermajority then there is a depth $d>0$ for which the depth-$d$ suffix of $B$, $\bar{B}(d)$, includes only $P$-blocks.
\end{proposition}
\begin{proof} [Proof of Proposition~\ref{proposition:equivocation-free-suffix}]
Consider a miner $q \in \Pi$ equivocating in $B$.  Since $B$ is disseminating, there is some $d$-suffix of $B$ in which all blocks observe these equivocating blocks. Let $B'$ be the  $d+1$-suffix of $B$, so that  every block point to some block that observes the equivocation by $q$. We claim that $B'$ cannot include any $q$ block.  Assume by way of contradiction that $b' \in B'$ is a $q$-block.  Since $B$ is cordial, $b'$  points to some block by a correct miner, which by assumption observes the equivocation by $p$.  Since $B$ is equivocation-repellent, $b' \notin B$.  A contradiction.  If we take the maximal such $d$ over all equivocators, then this $d$-suffix does not include blocks by any equivocator, namely it includes only $P$-blocks. 
\end{proof}

\begin{lemma}[Blocklace Common Core]\label{lemma:common-core}
	Let $B$ be a cordial blocklace,  $P \subset \Pi$ the set of miners that do not equivocate in $B$ and are disseminating in $B$, and $r>0$ a depth for which the depth-$(r-1)$ suffix of $B$, $\bar{B}(r-1)$, includes only $P$-blocks ($r$ exists by Proposition \ref{proposition:equivocation-free-suffix}), and hence equivocation free. If $P$ is a supermajority then there is a supermajority, of $r$-round blocks $\hat{B} \subset B$,  referred to as \temph{common core}, s.t. every $(r+3)$-round block approves all blocks in $\hat{B}$.
\end{lemma}

\begin{proof}[Proof of Lemma \ref{lemma:common-core}]
	Let $B$ and $r$ be as assumed by the Lemma, and let $P$ be the set of miners that have $(r+2)$-round blocks in $B$.  Since $B$ is cordial and disseminating it is infinite of depth $> r+3$ (Observation \ref{observation:infinite-dissemination}), it follows that $|P| \ge 2f+1$.
	Define a table $T$ with rows and columns indexed by $P$.
	Each $(r+2)$-round $p$-block of a miner $p \in P$ observes $(r+1)$-round blocks by at least $2f+1$ miners, which includes at least blocks by $f+1$ miners of $P$, represented in $T$.  
	For $p,q\in P$, entry $T[p,q]$ in the table is $1$ if the $(r+2)$-round $p$-block observes the $(r+1)$-round $q$-block, and $0$ otherwise.  Observe that if $1$ appears in  entry $T[p,q]$, the $(r+2)$-round $p$-block observes all the $2f+1$ $r$-round blocks observed by the $(r+1)$-round block of $q$.
	
	Since all miners in $P$ have $(r+2)$-round blocks,  $T$ contains  at least $(2f+1)(f+1)$ entries with $1$. This implies that there is a miner in $P$, say $\bar p$, that appears in at least $f+1$ rows; let $\bar{P}$ be the set of miners indexing $\bar p$ rows and $\bar b$ (blue dot in Fig. \ref{figure:common-core}) the $(r+1)$-round $\bar p$-block.  
	Thus, the $(r+2)$-round blocks of miners in $\bar{P}$ (thick green line at round $r+2$) observe $\bar b$. We argue that $[\bar{b}]$ is a common core.
	First, note that as $\bar b$ is cordial, $[\bar{b}]$ includes 
	$2f+1$ $r$-round blocks.  Second, consider any $(r+3)$-round block $b$ (green dot).  It observes $2f+1$ $(r+2)$-round blocks (thick red line), so it also observes at least one of the $(r+1)$-round blocks (black dot) of  the $f+1$ miners of $\bar{P}$, which in turn observes
	$\bar b$. Thus $[\bar{b}] \subseteq [b]$, with $[\bar{b}]$, satisfying the requirements of a common core.
\end{proof}

\begin{corollary}[Super-Ratified Common Core]\label{corollary:ratified-common-core}
	Under the same conditions as Lemma \ref{lemma:common-core} and assuming $\hat{B}$ is a common core, then every $(r+4)$-round block in $B$ ratifies every block in $\hat{B}$, Hence every member of the common core $\hat{B}$, is super-ratified in $B(r+4)$.
\end{corollary}

\begin{proof}[Proof of Corollary \ref{corollary:ratified-common-core}]
	Under the assumptions of the Corollary, let $\hat{B}$ be a common core and consider any $(r+4)$-block $b \in B$ (purple dot).
	Being cordial, $b$ observes $2f+1$ $(r+2)$-round blocks (thick purple line).
	By Lemma \ref{lemma:common-core}, each of these blocks observes each block in $\hat{B}$. Hence $b$ ratifies every block in $\hat{B}$.
\end{proof}

\begin{corollary}[Liveness of Common Core]\label{corollary:liveness_common_core}
	Let $B$ be a disseminating and cordial blocklace.  Then there is a $d>0$ for which the common core (Lemma \ref{lemma:common-core}) holds for $B$ and for any round $r\ge d$.
\end{corollary}
\begin{proof}[Proof of Corollary~\ref{corollary:liveness_common_core}]
According to Proposition \ref{proposition:equivocation-free-suffix}, $B$ has an equivocation-free suffix $B'$, to which
Lemma \ref{lemma:common-core} applies.
\end{proof}

\begin{proposition}[Leader-Liveness of Cordial Miners Asynchrony Protocol]\label{proposition:cordial-leader-liveness-a}
	The blocklace produced by a run of a Cordial Miners asynchrony protocol is leader-live with probability 1.
\end{proposition}

\begin{proof}[Proof of Proposition \ref{proposition:cordial-leader-liveness-a}]
	Let $B$ be the cordial blocklace produced by a run of a Cordial Miners asynchrony protocol, $P\subseteq \Pi$ the supermajority of miners correct in the run, and let $r>0$ be any round for which the $r-1$ suffix of $B$, $\bar{B}(r-1)$, is equivocation-free, $r \text{ mod } w = 0$, where $w= 5$ (Line \ref{alg:w-5}).
	According to Corollary \ref{corollary:ratified-common-core}, a supermajority $\hat{B}$ of the round-$r$ blocks is super-ratified by all round-$r+4$ blocks.  As the leader of round $r$ is selected at random, and retrospectively after the common core $\hat{B}$ has been established, the probability that the elected leader is super-ratified, and hence final, is at least $\frac{|P|}{n}$.  As this holds also for the leader of any round following $r$, the probability that for any depth $d\ge r$, a leader in $\bar{B}(d)$ has a final leader is 1, hence $B$ is leader-live with probability 1.
\end{proof}

\begin{proposition}[Cordial Miners Protocol Liveness]\label{proposition:CM-liveness}
	The Cordial Miners protocols for asynchrony and eventual synchrony are live.
\end{proposition}

\begin{proof}[Proof of Proposition \ref{proposition:CM-liveness}]
	According to Propositions  \ref{proposition:cordial-leader-liveness-es} and \ref{proposition:cordial-leader-liveness-a}, the blocklace produced in any computation of the Cordial Miners protocols for eventual synchrony and asynchrony is leader-live with probability 1. 
	According to Proposition \ref{proposition:tau-liveness}, if the function $\tau$ is applied to a sequence of blocklaces that converge to a leader-live blocklace $B$ then any $b \in B$ appears eventually in the output of $\tau$. If the blocklace $B$ is leader-live with probability 1 then any $b \in B$ appears eventually in the output of $\tau$ with probability 1,
	which is the liveness requirement of ordering consensus protocols (Def. \ref{definition:safety-liveness}). Hence the Cordial Miners protocols are live.
\end{proof}

\section{Future Direction and Optimizations} \label{appendix-section:futureDirection}

Several optimizations are possible to the protocol instances presented, which we intend to explore in future work:
\begin{itemize}
	\item As faulty miners are exposed, they are repelled and therefore need not be counted as parties to the agreement, which means that the number of remaining faulty miners, initially bounded by $f$, decreases.  As a result, the supermajority needed for finality is not  $\frac{n+f}{2n}$ (namely $2f+1$ votes in case $n = 3f+1$), but $\frac{n+f-2f'}{2(n-f')}$, where $f'$ is the number of exposed faulty miners, which converges to a simple majority ($\frac{1}{2}$) among the correct miners as more faulty miners are exposed and $f'$ tends to $f$.
	
	\item Once faulty miners are exposed and repelled, their slots as leaders could be taken by correct miners, improving the good cases and expected complexity.
	
	\item A hybrid protocol in the spirit of Bullshark~\cite{giridharan2022bullshark} can be explored. Such a protocol would employ two leaders per round -- deterministic and random,  try to achieve quick finality with the deterministic leader, and fall back to the randomly-selected leader if this attempt fails.
	
	\item Bullshark makes a key observation that fairness (namely, that every block of a correct miner eventually is output by $\tau$) and garbage collection cannot be achieved together in asynchronous networks.
	Rather, fairness can be achieved in synchronous periods, and therefore in the ES version of \sys one can achieve fairness after GST and also garbage collect old blocks in the blocklace.
	We plan to use similar techniques as Bullshark does for garbage collection.
	
	\item Exclusion of non-responsive miners: A miner $p$ need not be cordial to miner $q$ as long as $q$ has not observed a previous block $b$ sent to $q$ by $p$. If $q$ fail-stopped, then $p$ should definitely not waste resources on $q$; if $q$ is only suspended or delayed, then eventually it will send to $p$ a block observing $b$, following which $p$---being cordial---will send to $q$  the backlog $p$ has previously refrained from sending, and is not observed by the new block received from~$q$.
\end{itemize}

\end{document}